\documentclass[letterpaper,twocolumn,10pt]{article}
\usepackage{usenix2025_SOUPS}

\usepackage{tikz}
\usepackage{amsmath}
\PassOptionsToPackage{table,xcdraw}{xcolor} 

\usepackage{textgreek}
\usepackage{booktabs}
\usepackage{multirow}
\usepackage{subcaption}
\usepackage{xcolor,colortbl}
\definecolor{gray}{rgb}{0.1,0.1,0.1}
\usepackage[T1]{fontenc}
\usepackage[english]{babel}
\usepackage{graphicx}
\usepackage[flushleft]{threeparttable}
\usepackage{adjustbox}
\usepackage{subcaption}
\usepackage{longtable}
\usepackage{threeparttablex}

\usepackage{filecontents}

\usepackage{xspace}     
\usepackage{xpunctuate} 

\newcommand{\inlinequote}[1]{\emph{``#1''}\allowbreak} 

\newcommand{\blockquote}[2]{           
    \begin{quote}\emph{``#1''} (#2)\end{quote}\allowbreak}

\newcommand{\parHeading}[1]{\vspace{2mm}\noindent{\textbf{#1.}\allowbreak}}

\newcommand{\ie}[1]{(i.e., {#1}\allowbreak)}

\newcommand{\eg}[1]{(e.g., {#1}\allowbreak)}

\newcommand{\entry}[1]{P$_{e}#1$}

\newcommand{\diary}[1]{P$_{d}#1$}

\newcommand{\codesign}[1]{P$_{c}#1$} 

\usepackage{xargs}      
\usepackage{soul}       
\usepackage{color}      

\definecolor{LIGHTPINK}{RGB}{237,157,202}
\definecolor{LIGHTRED}{RGB}{210,121,121}
\definecolor{LIGHTORANGE}{RGB}{230,170,50}
\definecolor{LIGHTGOLD}{RGB}{210,194,121}
\definecolor{LIGHTGREEN}{RGB}{121,210,121}
\definecolor{LIGHTAQUA}{RGB}{121,206,210}
\definecolor{LIGHTBLUE}{RGB}{121,124,210}
\definecolor{LIGHTPURPLE}{RGB}{153,102,255}
\definecolor{RED}{RGB}{178,34,34}
\definecolor{GRAY}{RGB}{166,166,166}
\definecolor{WHITE}{RGB}{255,255,255}

\newcommandx{\jane}[2][1=] 
    {\setulcolor{LIGHTGREEN}{\ul{#1}} \textcolor{LIGHTGREEN}
    {[\textbf{Jane:} #2]}}
\newcommandx{\guest}[3][1=]
    {\setulcolor{LIGHTORANGE}{\ul{#1}} \textcolor{LIGHTORANGE} 
    {[\textbf{#2:} #3]}}

\begin{document}

\date{}

\title{\Large \bf Trust-Enabled Privacy:\\Social Media Designs to Support Adolescent User Boundary Regulation}

\def\plainauthor{Author name(s) for PDF metadata. Don't forget to anonymize for submission!}

\author{
{\rm JaeWon Kim}\\
University of Washington
\and
{\rm Robert Wolfe}\\
University of Washington
\and
{\rm Ramya Bhagirathi Subramanian}\\
Independent Contributor
\and
{\rm Mei-Hsuan Lee}\\
Independent Contributor
\and
{\rm Jessica Colnago}\\
Independent Contributor
\and
{\rm Alexis Hiniker}\\
University of Washington
} 

\maketitle

\pagenumbering{gobble}

\begin{abstract}
Adolescents heavily rely on social media to build and maintain close relationships, yet current platform designs often make self-disclosure feel risky or uncomfortable. Through a three-part study involving 19 teens aged 13–18, we identify key barriers to meaningful self-disclosure on social media. Our findings reveal that while these adolescents seek casual, frequent sharing to strengthen relationships, existing platform norms often discourage such interactions. Based on our co-design interview findings, we propose platform design ideas to foster a more dynamic and nuanced privacy experience for teen social media users. We then introduce \textbf{\textit{trust-enabled privacy}} as a framework that recognizes trust---whether building or eroding---as central to boundary regulation, and foregrounds the role of platform design in shaping the very norms and interaction patterns that influence how trust unfolds. When trust is supported, boundary regulation becomes more adaptive and empowering; when it erodes, users resort to self-censorship or disengagement. This work provides empirical insights and actionable guidelines for designing social media spaces where teens feel empowered to engage in meaningful relationship-building processes.
\end{abstract}

\section{Introduction}
Social media now serves as the foundation of social interactions for teenagers~\cite{Anderson2023-qk}. However, mainstream platforms like Instagram and TikTok have increasingly shifted their focus toward engagement with ``media'' rather than fostering truly ``social'' interactions~\cite{Prinstein2023-pd}. While these platforms excel at maintaining weak ties~\cite{Ellison2011-ld}, they are less effective at fostering the closer relationships crucial for youth well-being~\cite{kim2025socialmediafeellike}. Research suggests that social media may have contributed to a decline in the quality of social interactions~\cite{Putnam2000-ic, mcpherson2006social}, as it often neglects features that support the development of close interpersonal bonds~\cite{Lam2012-ct}. Teens seek authentic interactions online~\cite{10.1145/3686909, Reddy2024-ic}, yet self-disclosure---one of the most fundamental mechanisms for relational closeness---remains particularly difficult in digital spaces~\cite{Hogan2010-oh, Lenhart2015-cm}.

Privacy concerns play a significant role in shaping adolescent self-disclosure behaviors. Teens must navigate a complex landscape of both actual privacy risks and perceived threats amplified by alarmist narratives~\cite{Malkin-2022-RuntimePermissionsAssistants-p, kim2025privacysocialnormsystematically, de2020contextualizing}. This results in feelings of resignation, fear, and helplessness, discouraging meaningful social interactions. Despite advocacy for resilience-based approaches to privacy~\cite{Wisniewski-2018-PrivacyParadox-l, Agha2023-mu}, mainstream social media platforms and regulatory policies remain heavily prevention-focused~\cite{meta-sue, Other-Other-NYState2023-S7694A-p, Kim-2024-AustraliaBarred-y}, reinforcing a culture of anxiety rather than empowerment~\cite{Weinstein2022-rh}. 

Furthermore, privacy on social media is inherently networked\cite{marwick2014networked, Petronio2002-ce}. While platforms tend to emphasize individual privacy controls, teens' primary concerns often stem from interpersonal privacy risks---such as peer scrutiny, judgment, or parental oversight---rather than corporate data collection\cite{zhao2022understanding}. However, current privacy designs largely ignore the collective nature of boundary regulation, failing to support teens in navigating evolving trust dynamics with their social circles~\cite{Lowens-2025-MisalignmentsDemographicFacebook-d}.

Given the significance of supporting self-disclosure as a mechanism for meaningful social connection and recognizing the role of trust in adolescents' boundary regulation, we explore the following research questions:
\begin{itemize}
    \item \textbf{RQ1:} How do teens navigate self-disclosure within social media environments, and what factors influence their decisions to share?
    \item \textbf{RQ2:} How does social media design support or undermine trust-based self-disclosure among teens?
\end{itemize}
To answer these questions, we conducted a three-part study with 19 adolescents aged 13 to 18. The study comprised an entry interview to understand teens' concerns regarding online self-disclosure, a one-week diary study to capture specific instances where they hesitated to disclose meaningful content, and a co-design exit interview to explore solutions that would better support self-disclosure in digital spaces. 

Our findings reveal that teens desire a social media environment where they feel comfortable sharing everyday moments with their online friends as a way to maintain and strengthen relationships. However, while self-disclosure is more intuitive within trusted circles, interactions with audiences whose trust levels are ambiguous often lead to defaulting to disengagement. Teens recognize that trust can develop over time, yet platform designs often fail to support this progression. Instead, interactions with not-yet-trusted ``Friends'' are shaped by platform norms that encourage shallow engagement and, in many cases, erode rather than build trust. The teen participants suggested several features to counter this by allowing greater alignment with audience expectations, facilitating selective and context-specific sharing, and reducing the (perceived) social risk of disclosure.

Through this work, we present empirical evidence that teens want to engage in self-disclosure but feel constrained by current social media design limitations. We highlight specific interaction patterns that undermine trust and propose design affordances---both existing and new---that can shift these interactions toward trust-building. Additionally, we introduce design guidelines that reframe privacy not as a trade-off between control and disclosure but as a dynamic, \textbf{\textit{trust-enabled}} process that fosters relational development. We encourage social media designers to move beyond traditional information-control-based privacy approaches and consider self-disclosure as a relational practice that requires both initial trust scaffolding and mechanisms for reinforcing trust through repeated interactions. By integrating these insights, platforms can better support teens' social needs, ultimately contributing to a more empowering and trust-enabling online environment.

\section{Background and Related Work}
Understanding the intersection of trust, privacy, and social media design is crucial for developing platforms that effectively support adolescent users in regulating their boundaries. Existing research highlights the limitations of current social media privacy models and underscores the importance of trust in shaping adolescent disclosure and privacy behaviors. This section explores the literature on the kind of privacy adolescents seek, reviews existing social media designs, and examines the role of trust in privacy management.

For our study, we focus on strengthening existing ties \eg{developing a closer relationship with an acquaintance added as a Follower on Instagram}, especially those ties that extend into real-life relationships~\cite{Nguyen2012-cc}. We also draw upon Altman's Social Penetration Theory~\cite{altman1973social} and view relationship-building from the perspective of repeated (reciprocal) exchanges of increasingly deepening and broadening self-disclosure. We take the definition of self-disclosure from Derlega et al.~\cite{derlega1993self}, i.e., \textit{``a deliberate or voluntary activity whereby people reveal information, thoughts, and feelings about themselves to at least one other person during an interaction.''} 

We focus on platforms where people share content with multiple others (1:N) with some degree of permanence, a focus that emerged naturally from our interviews rather than one-to-one messaging or ephemeral group chats. We refer to these as \textit{broadcast social media} for convenience. In these platforms, challenges around self-disclosure are more prevalent and harder to address, partly due to context collapse. 

We also scope trust to \textbf{\textit{interpersonal trust}}---the belief that others have one's best interests at heart---aligning with relational trustworthiness factors such as benevolence and integrity~\cite{Nguyen2009-ri}. Within the scope of teen social media privacy that we examine in this paper, trust manifests as confidence that viewers will not misuse disclosed information and will respond to or judge what is shared with skepticism.

Lastly, we acknowledge that throughout the paper, we alternate between specifying teen users and users in general.This is because---as per the Universal Design principles and much literature in privacy where findings in adults extend to teens and vice versa such as context collapse~\footnote{We also confirm that the concept of Privacy as Trust~\cite{Waldman2018-dc}, the main framework we apply in the current paper, extends to teens in Section~\ref{section:4-1}}---\textbf{\textit{much of what is relevant to teens is also applicable to the general population}}~\cite{Christofides-2012-HeyMomAdults-o, burgstahler2010universal}.

\subsection{Teens Privacy Concerns and Awareness}
Adolescents' developmental characteristics make online sharing particularly high-stakes. They experience heightened significance of peer acceptance and sensitivity to social rejection~\cite{Somerville2013-zf}; to maintain social status, teens often expand their ``Friend'' networks beyond the circle of friends they trust~\cite{Yau2019-ab}, intensifying context collapse challenges. Real-world social pressures further complicate privacy management~\cite{Weinstein2022-rh}, as peers' digital documentation and phenomena like \inlinequote{cancel culture} \ie{the practice of publicly shaming and boycotting individuals for perceived wrongdoings} and \inlinequote{receipts} \ie{proof screenshots, screen recordings, or messages saved for evidence of an indiscretion} leave teens feeling vulnerable and disempowered about their privacy~\cite{Weinstein2022-rh, BoydDanah2014ICTS}. 

Further, given context collapse~\cite{BoydDanah2014ICTS}, users often feel compelled to share only the \inlinequote{lowest common denominator}~\cite{Hogan2010-oh}, content deemed universally appropriate. The hyperpersonal nature of online communication~\cite{walther2011introduction} often leads to exaggerated or inaccurate understandings of individuals based on limited cues, which makes curated and infrequent posting more high-stakes. While these challenges affect people across all demographics, they are particularly relevant to teens, whose developmental characteristics tend to exacerbate these issues.

Contrary to the common perception that teenagers are indifferent to or oblivious of online privacy risks, research consistently demonstrates that they are acutely aware of the potential dangers of disclosure online and actively engage in privacy management~\cite{Cho-2018-CollectivePrivacyValidation-h} employing various tactics such as social steganography---embedding hidden meanings in posts---or manually filtering audiences~\cite{DeWolf-2014-ManagingPrivacyFacebook-c}. They recognize that online disclosure can lead to unforeseen consequences, from various cyber risks like cyberstalking and identity theft to exposure caused by people they know~\cite{Agha2023-mu, zhao2022understanding}. Teens also understand that once information is shared online, it is difficult to control its spread given the networked nature of social media platforms~\cite{Dienlin-2015-PrivacyParadoxBehaviors-w, DeWolf-2014-ManagingPrivacyFacebook-c}.

However, such an awareness does not eliminate their anxiety~\cite{kim2025privacysocialnormsystematically}. Many teen users experience stress and uncertainty about their digital footprint, fearing unintended consequences of their disclosures~\cite{Williams2023-jm}. They understand that privacy is not solely an individual concern but is shaped by the actions of others in their social networks~\cite{marwick2014networked}. Given the risks of privacy breaches that can occur even with careful disclosure, teens often experience resigned pragmatism---engaging in privacy protection measures while simultaneously feeling a sense of resignation~\cite{Acquisti2017-ez, Weinstein2022-rh}. In worse cases, this can lead to outright privacy resignation~\cite{Malkin-2022-RuntimePermissionsAssistants-p}, dysfunctional fear that affects quality of life~\cite{kim2025privacysocialnormsystematically}, and fatalistic \inlinequote{network defeatism}~\cite{de2020contextualizing}. Recognizing the contextual and relational nature of their privacy~\cite{zhao2022understanding}, teens desire to have more agency and feel more empowered in managing their online presence~\cite{kim2025privacysocialnormsystematically, kim2025socialmediafeellike}.

\subsection{Privacy as Dynamic Boundary Regulation}
Privacy in social media, especially for teens, is not solely about restricting access to personal information. It is an ongoing process of interpersonal boundary regulation~\cite{Page-2019-PragmaticToolFeatures-p, Barkhuus-2012-MismeasurementPrivacyHCI-d, DeWolf-2014-ManagingPrivacyFacebook-c}. However, instead of supporting adolescents in dynamically managing their boundaries in ways that align with their social needs, social media privacy narratives and platform designs often frame privacy as a trade-off---forcing users to relinquish control in order to engage socially~\cite{Wisniewski-2015-GiveSocialWant-z, Wisniewski-2016-FramingMeasuringUsers-w}. Further, social media safety measures often prioritize surveillance and prevention-based approaches over resilience-based ones~\cite{Wisniewski-2018-PrivacyParadox-l}---for example, privacy-preserving tools that encourage open dialogue~\cite{Chouhan-2019-Co-designingCommunityTogether-x}. Similarly, the concept of group privacy~\cite{Choksi-2024-PrivacyGroupsMatters-b} is often neglected, often emphasizing individual control rather than collective boundary-setting mechanisms. Such views and designs that frame self-disclosure as inherently unsafe largely ignore the social benefits of self-disclosure that are especially critical to adolescents~\cite{Davis2012-bq, Weinstein2022-rh}. 

Privacy is also context- and norm-dependent~\cite{Nissenbaum-2004-PrivacyContextualIntegrity-j}. It is dynamic and situational, influenced by factors such as the intended audience and perceived long-term potential risks~\cite{Barkhuus-2012-MismeasurementPrivacyHCI-d}. Furthermore, privacy preferences shift over time~\cite{Barkhuus-2012-MismeasurementPrivacyHCI-d}, necessitating adaptable privacy tools that reflect evolving needs. However, there is often a mismatch between users' privacy expectations and needs versus the actual control they have over the data they disclose~\cite{Wisniewski-2015-GiveSocialWant-z}. Default privacy settings often fail to provide the meaningful protection necessary while navigating the complex and nuanced \textit{contexts}~\cite{Acquisti2017-ez}. Structural challenges in managing audience and disclosure are well captured by Knijnenburg et al.'s Network and Territorial Boundaries~\cite{knijnenburg2022modern}, which highlight how overlapping audiences and persistent content complicate boundary regulation. This shapes platform norms that are not conducive to co-owners of boundaries upholding their privacy expectations~\cite{Choksi-2024-PrivacyGroupsMatters-b, Abaquita-2020-PrivacyNormsIntegrity-m}. 

Compounding this issue, the often-found one-size-fits-all approach to privacy settings does not account for the diverse needs of users---especially those of adolescents---who navigate varying levels of social exposure and relationships~\cite{Wisniewski-2015-GiveSocialWant-z}. In addition, many platforms prioritize public interaction over privacy, making it difficult for users to navigate context collapse and manage audience segmentation effectively~\cite{10.1145/3686909, Yau2019-ab}. The interplay of these hurdles results in teens expressing concerns about privacy but still engaging in self-disclosure due to social benefits or peer influence~\cite{kim2025privacysocialnormsystematically, Wisniewski-2018-PrivacyParadox-l}. 

\subsection{Effective Privacy Management via Trust}
Trust and boundary regulation are deeply interconnected~\cite{derlega1979self}. Communication Privacy Management (CPM) theory~\cite{Petronio2002-ce}, for example, highlights how privacy rules and co-ownership of information depend on trust. Trust is fundamental in an environment where users rely on each other to uphold privacy norms~\cite{Waldman2016-em, Waldman2018-dc}. A good example of this is trust in \inlinequote{Bounded Social Media Places} (BSMPs)~\cite{Malhotra-2024-"whatPostPlaces-u}, such as private groups or direct messages, where users assume their information remains within a limited audience. Empirical research has also demonstrated that the relationship between privacy concerns and self-disclosure is significantly mediated by trust~\cite{Joinson2010-cf}. While individuals with higher privacy concerns may be less likely to disclose information, this effect is substantially reduced when they trust the recipient of the information.

Conversely, breaches of trust---whether through breaches, unintended exposure, or platform failures---create \inlinequote{boundary turbulence}~\cite{Petronio2002-ce}, leading to reduced self-disclosure, blocking, and stricter regulations. However, current social media platforms often fail to reinforce privacy responsibilities with clear expectations or reliable controls~\cite{Choksi-2024-PrivacyGroupsMatters-b, Paci-2019-SurveyAccessSystems-a}. Moreover, social influence plays a crucial role in shaping privacy behaviors---users are significantly more likely to adopt security features when they see their peers doing the same~\cite{Akter-2023-EvaluatingImpactSecurity-h}. However, most platforms fail to harness these collective privacy dynamics, instead burdening individual users with the full responsibility of managing their privacy~\cite{Schnitzler-2020-SoKManagingData-h, Humbert-2019-SurveyInterdependentPrivacy-f, hargittai2016can, kim2025privacysocialnormsystematically}. This lack of support not only increases the complexity of privacy management but also exacerbates the pervasive negativity in social media~\cite{kim2025privacysocialnormsystematically}.

\subsection{Existing Design Approaches} 
To address these issues around boundary regulation in social media, usable privacy researchers have proposed several design interventions to enhance trust and privacy while accommodating the social needs of users. The overarching theme around the spectrum of ideas is the recognition that privacy is not merely about limiting access but rather about enabling users to negotiate social boundaries dynamically. 

One approach is community-based privacy management mechanisms~\cite{Akter-2023-EvaluatingImpactSecurity-h}. These designs acknowledge that users co-manage information and establish shared privacy norms with their friends. However, existing platforms often lack sufficient support for collaborative privacy management mechanisms~\cite{Cho-2018-CollectivePrivacyValidation-h}. To address this gap, tools were designed to enable users to negotiate privacy boundaries together, such as features that allow co-owners of a post to jointly decide its audience or require mutual approval for tags and shares~\cite{Humbert-2019-SurveyInterdependentPrivacy-f, Wisniewski-2016-FramingMeasuringUsers-w}. Another important approach is \inlinequote{community oversight}~\cite{Akter-2024-ExaminingCaregivingSecurity-n, Akter-2023-EvaluatingImpactSecurity-h}, a mechanism that leverages trusted social groups to help individuals manage their digital safety regarding mobile app use. The approach is based on the principles of transparency, making community members' app installations and permissions visible, and awareness, keeping members informed of each other's practices and changes. It is intended to build trust and foster a sense of shared responsibility for privacy and security within the group.

Another key design principle is based on the contextual integrity framework~\cite{Nissenbaum-2004-PrivacyContextualIntegrity-j} where system design would align with users' expectations of information flow. One example is adaptive privacy defaults, which dynamically adjust visibility settings based on contexts such as content sensitivity, audience type, or prior user preferences~\cite{Acquisti2017-ez, Wisniewski-2015-GiveSocialWant-z}. For example, platforms could suggest context-sensitive audience suggestions recommending different levels of visibility for casual updates versus personal reflections~\cite{Mondal-2019-MovingBeyondMedia-p, Sleeper-2013-PostWasntFacebook-m}. Another approach is \inlinequote{role- and relationship-based access control} (RBAC \& ReBAC)~\cite{Paci-2019-SurveyAccessSystems-a}, where access is determined not just by predefined categories (e.g., ``Friends'' vs. ``Public'') but by the relational context between the user and their audience. Real-time audience feedback---such as a preview of who will see a post---could help users make more informed decisions before sharing~\cite{Acquisti2017-ez, Paci-2019-SurveyAccessSystems-a}. Further, temporary access controls could let users set expiration dates on posts or limit visibility based on contextual factors, like whether a viewer has interacted with them recently~\cite{Ayalon-2013-RetrospectivePrivacyNetworks-t, Schnitzler-2020-SoKManagingData-h, Mondal-2019-MovingBeyondMedia-p}. These design ideas enable users to regulate disclosure dynamically in ways that better align with their evolving social contexts.

Lastly, platforms could move beyond one-size-fits-all approaches to accommodate different attitudes toward boundary regulation~\cite{Wisniewski-2015-GiveSocialWant-z, Stutzman-2012-BoundaryRegulationMedia-n}. Given that privacy needs vary widely across users---some prioritize risk management, while others may be more concerned with relational dynamics~\cite{Page-2019-PragmaticToolFeatures-p, Wisniewski-2015-GiveSocialWant-z}---platforms could offer customizable privacy models, letting users choose different levels of privacy strictness depending on their current preferences~\cite{Colnago-2023-ThereReverseBehaviors-j}. For example, discretionary privacy features like a ``hide'' option (instead of blocking or unfriending) could reduce social friction~\cite{Page-2019-PragmaticToolFeatures-p}. Nudging mechanisms, such as gentle reminders to review audience settings, can encourage safer practices without disrupting user autonomy~\cite{Acquisti2017-ez, Agha2023-dv, Agha2023-mu}. Namara et al.~\cite{Namara-2022-EffectivenessAdaptationSites-y} compare adaptation methods for tailoring privacy settings, showing that strategies like contextual prompts and user-driven customization can improve alignment with user expectations. By recognizing and accommodating these individual differences, social media systems can provide more flexible, user-centered privacy solutions that support both nuanced relational concerns and pragmatic risk avoidance.
\section{Method}
The study consisted of three procedures: 1) a 30-minute entry interview, 2) daily reflection surveys throughout 7 days (with a minimum requirement of 5 or more entries), and 3) a 60-minute exit interview involving co-design activities. All interviews were conducted individually.


\subsection{Materials and Procedures}
\vspace{2mm}
\noindent\textbf{Screening survey.}
We first administered a screening survey prior to the interviews. The screening survey included basic demographic questions, three open-ended questions about: concerns or reservations about sharing on social media, existing features that currently facilitate sharing, and suggestions for features that could make sharing easier. The survey also contained 26 questions from the peer attachment scale \cite{Armsden1987-ai}.

\vspace{2mm}
\noindent\textbf{Entry interview.} We introduced the overall purpose and procedure of the study. We also asked about social media platforms where participants felt most at ease sharing posts, what they shared, who they shared with, any concerns they have when sharing, and any specific features or designs that affect the comfort of sharing.

\vspace{2mm}
\noindent\textbf{Diary study.} The diary study captured data in three scenarios: 1) participants shared something on social media, 2) they considered sharing but did not share, and 3) they neither considered nor shared anything. In the first scenario, we asked participants about the content and location of their sharing, the audience, their comfort level, and any features that increased their comfort. We also asked about any risks or benefits they associated with the sharing. For the other two scenarios, we posed similar questions focused on what they had intended or might have shared for the sake of building relationships. Additionally, we asked why they chose not to share and what features could have made them more comfortable doing so. The full questionnaire is available in the supplementary materials. We required at least five diary entries over the seven-day period. Most participants submitted five diary entries, with some submitting between six and 19 entries.

\vspace{2mm}
\noindent\textbf{Co-design interview.} We used a virtual whiteboard, created with the Miro platform \cite{miro}, to anchor a conversation about the barriers participants face to ``sharing comfortably on social media.'' We presented nine sticky notes\footnote{In no particular order, 1) prevent getting ignored; 2) be able to control who gets to see what you share; 3) make sharing easier, more casual, more spontaneous, less pressure; 4) knowing what the right amount of sharing is; 5) defining and communicating privacy rules/norms; 6) encourage more effortful communication/response from friends; 7) prevent negative reactions \eg{judgment, misunderstanding}; 8) increase trust with friends (followers); 9) prevent posts being shared out of context} that encapsulated the major concerns that participants shared in the screening survey and during the entry interviews. We also included additional subcategories related to each of the nine overarching concerns. Using virtual sticky notes, participants listed their concerns in order of personal relevance, and we structured our interviews based on that order. We addressed each sticky note in the sequence they considered most important. For each sticky note, we asked participants to explain the meaning and relevance of each concern and to suggest potential design solutions.

One member of the research team conducted the semi-structured entry and co-design interviews via Zoom, which lasted 30 to 40 minutes and 60 to 90 minutes, respectively. Participants received \$10 and \$20 Amazon gift cards for the entry and co-design interviews. All participants received a \$25 Amazon gift card for their diary study participation, regardless of the number of entries they submitted. Those who completed the full co-design study received an additional \$5 gift card.

\subsection{Participants and Recruitment}
We invited 80 participants from an established participant pool---composed of individuals who had taken part in one of our previous studies or expressed an interest in future research via other studies---to complete the screener survey for the co-design interview. The participant pool consisted of teens aged 13–18 across the U.S., recruited through an Instagram ad targeting specifically the age group and location. We selected 22 of the 47 teens, chosen based on the thoroughness of their responses to the open-ended questions and the diversity of their demographic data and peer attachment responses. Of the 22 invitees, 20 enrolled in the study, one withdrew following the entry interview, and 19 completed all three parts of the study.

Twelve participants self-identified as girls, five as boys, and two as non-binary. Their age ranged from 13 to 19, with a mean of 15.8 years (SD=1.7). Ten identified as White, six Asian, two as Black or African American, and one as White and Asian. Only three identified as Hispanic or of Latin-American origin. Participants' social media usage was diverse but with a significant predominance of Instagram (19) and BeReal (16). These were also the platforms that participants reported sharing most frequently (8 and 6 participants, respectively) and more comfortably (7 and 5, respectively).
\footnote{See Appendix \ref{ref:individual-demographics} for individual demographic data}

\subsection{Reflexive Thematic Analysis}
The interviews were transcribed and lightly edited for clarity and readability while preserving the original meaning and intent of the participants' responses. We conducted a reflexive thematic analysis~\cite{braun2019reflecting} of the transcripts, ensuring a structured yet flexible approach to analyzing our data in a nuanced and exploratory manner not tied to any one existing framework.

Our analysis process began with three members of the research team independently analyzing the transcripts line by line using Google Docs~\cite{google-docs}. Following this initial coding phase, we convened for weekly discussion sessions to develop an initial set of codes based on our individual analyses. We then proceeded to inductively code the transcripts using Atlas.ti~\cite{atlas}, referring to the established codes while remaining open to identifying new themes. Through collaborative discussions, we reconciled identified codes, resolved disagreements, and progressively refined our themes through several iterations. This collaborative process continued until the second and last authors had reviewed over half of the transcripts (including both entry and co-design interviews), at which point we reached thematic saturation~\cite{saunders2018saturation}. After that, the first author independently coded the remaining transcripts using the established codes. 

Based on the themes that we identified during the thematic analysis, we developed high-fidelity prototypes of each of the eight design ideas that we co-designed with the teen participants. The design ideas and descriptions of the high-fidelity mock-ups of possible implementation of the design ideas are available in Table \ref{tab:designs} and prototypes in Appendix \ref{fig:prototypes}. These are discussed further in Section~\ref{section:4}.

All interview and diary study protocols and the final set of codes are available at \href{https://osf.io/ga3yr/}{https://osf.io/ga3yr/} via the Open Science Framework~\cite{Other-Other-Osf-n}.

\subsection{Ethical Considerations}
All procedures were approved by our Institutional Review Board (IRB) prior to data collection. Before the interviews, we obtained written consent from participants and parents. At the start of each interview, we reviewed the key points from the consent form and procedures, ensuring participants had the opportunity to ask any questions. Participants were informed of their right to decline questions, turn off cameras, and withdraw from the study without consequence. We emphasized our goal was to understand, not judge, their experiences. We then obtained verbal consent to proceed with the study and to record the interviews.
\section{Results}
\label{section:4}
Through our interviews and diary study data, we find that teens perceive casual, mundane, and frequent sharing of their daily lives as meaningful self-disclosure---helpful in building relationships but not too intimate or high-stakes. However, many hesitated to share due to concerns about their content being perceived as underwhelming or trivial. Our study also reveals that trust is crucial in how teens regulate boundaries around self-disclosure. When trust in a relationship feels uncertain, teens face a dilemma: self-disclosure is key to strengthening connections, yet within the social media environment, sharing with an ambiguous audience often feels more risky than rewarding. As a result, many teens default to self-censorship, limiting the potential for relationship building.

This dynamic suggests that the way social media privacy is portrayed may discourage trust-building rather than support it. We explore two overarching themes of barriers to meaningful self-disclosure: uncertainties in communication \ie{\textit{communication fog}} and a high-stakes environment that tends to skew these uncertainties towards negative interpretations \ie{\textit{low-grace culture}}. Additionally, we examine teens' design suggestions for addressing these barriers, providing insights into potential platform improvements. We denote quotes with (\entry{XX}) for entry interview data, (\diary{XX}) for diary entries, and (\codesign{XX}) for co-design interview data.

\subsection{Trust as Key for Boundary Regulation}
\label{section:4-1}
Through our interviews and diary studies, we found that trust functions as a key determinant in shaping teens' self-disclosure practices and their ability to regulate boundaries online---trusted friends or interactions that build trust facilitated more open and frequent sharing, while non-trusted individuals or \textit{ambiguous} connections led to hesitations.

\subsubsection{\textbf{Self-Disclosure as a Balancing Act Between Relational Benefits and Privacy Concerns}}
\label{section:4-1-1}
Participants recognized the relational benefits of disclosure. They desired to share mundane details or small updates from their lives on social media but hesitated due to concerns about how their content would be perceived. For example, one participant mentioned wanting to share small updates in their lives, such as \inlinequote{I made some microwave popcorn and at the end there were only 11 pop kernels at the bottom of the bag} (\entry{19}). One participant who initially had concerns about being perceived as oversharing, reflected: 

\blockquote{When\ldots{} I see something like this that maybe shares details that are a little more personal I definitely feel like I connect to it more, and that's when I kind of realize maybe like sharing a couple like personal details what might be considered oversharing to some people wasn't always a bad thing.}{{\codesign{08}}}

However, many teens refrained from posting such content due to an internalized expectation that self-disclosure must be curated. They worried that their posts were \inlinequote{not memorable and significant enough to be shared} (\diary{03}) or \inlinequote{not Instagram worthy} (\diary{08}), leading to a tendency toward self-censorship. Some participants expressed frustration at this pressure, wanting to \inlinequote{just post} (\codesign{09}) without feeling \inlinequote{judged} (\codesign{09}).

\subsubsection{\textbf{Trusted Friends as a Safe Space for Self-Disclosure}}
\label{section:4-1-2}
Teens shared that having a way to carve out their trusted circle of friends supported disclosure. As \diary{16} noted, \inlinequote{The limited nature of my friends on the app helped me feel comfortable, because I trust everyone on there.} Features such as Instagram's ``Close Friends'' reinforced this sense of security: \inlinequote{I shared these on my Close Friends story. These are people I am comfortable talking to at any time without it feeling weird or forced} (\diary{02}). Similarly, \diary{17} noted, \inlinequote{I always post a daily BeReal. I just love the app and how exclusive it is.} Moreover, a trusted circle alleviated concerns about judgment. \diary{16} elaborated, \inlinequote{These are people who either will not judge me or I simply do not care if they do.} \diary{19} echoed this sentiment that with close friends, they do not mind their posts being perceived as \inlinequote{not\ldots{} funny} or even \inlinequote{weird.}

\subsubsection{\textbf{Non-Trusted ``Friends'' as a Barrier to Self-Disclosure}}
\label{section:4-1-3}
The presence of ambiguous or distrusted connections, on the other hand, hindered self-disclosure. Teens frequently described situations where they wanted to share but refrained due to the potential presence of audiences they would not trust. On Instagram, for example, even with a private account, once someone is accepted as a ``follower,'' they gain immediate access to all shared content unless the user curates a ``Close Friends'' list. As \diary{16} explained, \inlinequote{Well I WANTED to share it with friends, but my followers on Instagram encompass more than that. I don't want to be judged.}

Platform norms further complicated boundary management. Many teens felt compelled to accept ``Follow'' requests, even from those who they did not fully trust, to avoid appearing rude. As \codesign{17} described, Instagram often resembled a \inlinequote{popularity contest} where the \inlinequote{follower to following ratio} (\codesign{15}) dictated social standing. This dynamic pressured users to \inlinequote{accept all the requests\ldots{} just to increase [their] follower count} (\codesign{05}) and made it harder to regulate self-disclosure. In contrast, platforms like BeReal, which de-emphasize follower metrics and provide clearer expectations for social boundaries, were seen as more conducive to trust-based boundary regulation: \inlinequote{The hidden follower count helps, so I don't feel pressured to get a lot of followers} (\diary{16}).

\subsubsection{\textbf{Evolving Boundaries and the Role of Trust}}
\label{section:4-1-4}
Our findings show that trust-building on social media mirrors that from prior studies: trust initiation begins with sharing, and repeated reciprocated interactions are essential for trust to develop and solidify. Participants highlighted that building trust with peers online starts with self-disclosure:
\blockquote{For me, building trust looks\ldots{} kind of like getting to post what you like, and maybe having a small two-minute conversation can help build a little more trust.}{\codesign{13}}
Similarly, \codesign{01} shared that trust is fostered through \inlinequote{messaging them and liking or commenting on each other's posts.} 

Participants also confirmed that favorable responses to self-disclosure are critical in trust-building. For example, \codesign{10} noted that trust comes mostly from \inlinequote{the way a person responds to a post that [they] send out.} They elaborated, \inlinequote{if I post a status update and the person maybe relates or comforts me or something, then that can obviously make me trust them.} Reciprocity was seen as key to building trust:
\blockquote{[Trust is a] mutual thing like if you reply more to me and I will talk to you more that trust kind of builds because we get to know each other more.}{\codesign{13}}
The role of effort was also emphasized, as comments that are \inlinequote{sweet, engaging, and creative,} for example, nurture trust and a sense of community (\codesign{06}, \codesign{16}). 

Conversely, a lack of reciprocity often leads to the erosion of trust. For example, \codesign{10} shared, \inlinequote{If they respond to it negatively, then my trust in them definitely decreases.} Non-responsiveness also damages trust, as \codesign{10} explained: if friends or followers \inlinequote{view it and not do anything} when they \inlinequote{explicitly ask for} something, it decreases trust. This lack of engagement, described as giving the \inlinequote{cold shoulder} (\entry{17}), discourages further self-disclosure (\codesign{01}, \entry{17}). Sometimes, trust erosion extends to groups, as \codesign{14} explained: if they \inlinequote{get like 300 views on the story but then only 150 likes}, it often signifies that \inlinequote{people saw it but didn't really do anything,} leading them to \inlinequote{take it the wrong way} and believe that people \inlinequote{don't like [them] anymore.}

\subsection{Communication Fog: Barrier to Boundary Regulation and Trust Calibration}
\label{section:4-2}
One key aspect that complicated trust-based boundary regulation was the uncertainty about how their posts are received on social media. Participants were unsure of platform norms, leaving them burdened with individual responsibility to navigate about what or how much to share. They also question whether their audience might judge them or not be interested in what they post. Additionally, interpreting reactions---especially low-effort, obligatory responses such as ``Likes''---is confusing, making it difficult to gauge genuine engagement or appreciation. These ambiguities create a ``communication fog'' that complicates trust calibration and boundary regulation, making users hesitant to share in ways that support relationship-building.

\subsubsection{Ambiguous Norms: Unclear Sharing Expectations and Individual Burden}
\label{section:4-2-1}
Participants highlighted the significant influence of peer norms on their self-disclosure behaviors, emphasizing their tendency to align their sharing with the perceived expectations within their social circles. Teens carefully calibrate their disclosures to neither exceed nor fall below the perceived norm, as \entry{02} noted: \inlinequote{Oftentimes, I kind of match what my friends post in terms of how public they are.} This balancing act reflects an ongoing negotiation of social expectations, with \codesign{07} stating: \inlinequote{So you don't want to overshare and you don't want to undershare.} 

However, many platforms lack clearly defined norms for what constitutes an appropriate level of sharing, leaving users to navigate ambiguous expectations independently. Participants interpret implicit norms:
\blockquote{[On Instagram,] Not famous people, in my experience, do not normally post random stuff like this, so it would be weird to make it my first post.}{\diary{16}}
Therefore, Instagram necessitates users to \inlinequote{make an active choice} (\codesign{05}) about when and what to share. Hence, self-initiated posting can sometimes be interpreted as \inlinequote{attention-seeking} (\diary{16}), and teens fear (potential) backlash. For instance, \codesign{12} had been asked \inlinequote{why do you post so much on Instagram?}, and \codesign{11} described having been ``unfollowed'' by a peer who deemed their content uninteresting: \inlinequote{He was like, oh you don't post anything interesting. All you do is post selfies.}

Participants expressed a desire for \textbf{guided disclosure}, where platforms would offer clear, explicit cues that establish expectations for how much, how often, and when to share. By creating shared reference points for appropriate disclosure, these prompts would reduce the burden of individual decision-making and reinforce the reciprocity of self-disclosure, reassuring users that their vulnerability is not being interpreted as an isolated attempt for attention but rather as a socially expected practice. 

More specifically, participants envisioned being able to make \inlinequote{a quick update on [their] life, even if it's quite boring} (\diary{05}). Platforms like BeReal, which enforce a collective norm of casual sharing at a predefined time, were appreciated for \inlinequote{give[ing] everyone a chance to shout out} (\codesign{05}). Almost all participants advocated for system-generated \textit{prompts} or reminders that would encourage low-stakes sharing and provide implicit permission to disclose, fostering a \inlinequote{sense of community} (\codesign{12}, \codesign{16}) while reducing concerns about \inlinequote{sharing too much} (\codesign{19}). Such interventions would also alleviate uncertainty when users feel like they \inlinequote{can't really figure out what to post} (\codesign{07}) by \inlinequote{giv[ing] ideas about what to share} (\diary{08}), making self-disclosure a more collectively guided and less individually scrutinized process.

\subsubsection{Ambiguous Loyalty: When Followers Feel Like Skeptics}
\label{section:4-2-2}
Participants expressed that social media ``friends'' or ``followers'' often lack the established trust of real-world friendships. While these connections did not warrant removal---sometimes due to fears of social repercussions or backlash---their intentions and judgments remained uncertain. This ambiguity created a challenge in boundary regulation as users struggled to assess how their audience might interpret their posts. As \diary{08} explains: \inlinequote{I don't want some of my friends to think I'm oversharing or being `cringey.' There are always risks. You never can really know who's viewing your content or who's REALLY following you.}

To mitigate these uncertainties and foster a greater sense of trust-based boundary control, participants proposed supporting \textbf{mutual commitment} in content consumption. One suggested mechanism involved posters sending invitations, requiring viewers to actively opt-in (\codesign{16}) or out (\codesign{03}): \inlinequote{A pop-up of like `Oh you've been invited to join'\ldots with a yes or a no button,} and offering \inlinequote{the option to either join it or leave if you don't want to follow those rules} (\codesign{11}). This approach redistributes accountability---allowing viewers to regulate their access while relieving sharers of sole responsibility for audience reactions. A participant explained how mutually agreed sharing, such as that on Snapchat private Stories, provides reassurance: 
\blockquote{If they joined that story, [then] they're kind of the ones subjecting themselves to it. If they didn't want to see it, they'd never had to join it.}{\entry{12}}

\subsubsection{Ambiguous Relevance: Struggles to Identify the Right Audience}
\label{section:4-2-3}
Participants expressed a strong desire to curate their audience, even when viewers were not overtly toxic or judgmental. This need stems from their multifaceted identities and diverse interests, which they prefer to compartmentalize for different social groups. Participants shared various interests that they felt would only engage a subset of their audience, from poetry (\entry{11}) to K-Pop (\entry{03}). One participant explained the challenge of this:
\blockquote{sometimes I don't want to bore some of my friends with [a specific interest of theirs]\ldots{} I don't really feel the need to share it with them because as much as I would like them to share my interest in it, I don't want them to find that kind of burdensome.}{\codesign{14}} 

While many platforms employ algorithms to curate content based on inferred interests, participants found these coarse and ineffective. Rather, participants advocated for self-managed and dynamic calibration of social boundaries through \textbf{contextual disclosure}, allowing them to strategically engage with groups based on the established and/or perceived potential for trust. Some users created elaborate systems of nested groups via Snapchat's private stories feature: 
\blockquote{I have four [private stories], each getting smaller to accommodate things I choose to share with\ldots{} The closer you are to me, the more (and smaller) private stories you're in.}{\diary{15}} 
\entry{16} reflected on the flexibility of Discord as providing similar benefits. 
These user-curated spaces would not just serve as secure places for existing trust but as controlled environments where users could gradually calibrate their social boundaries, engaging in disclosure that fosters emerging trust while minimizing the risks of sharing too broadly.

Participants were particularly keen on interest-based spaces, where shared interests would serve as a bridge to potential trust development. They noted that while common interests alone do not equate to trust, they provide an initial context that reduces the uncertainties of broader public sharing and, ultimately, allows trust to form over time. As \codesign{08} explained, such intentional segmentation \inlinequote{could definitely help people connect more because they'd feel like they had more to talk about.}

\subsubsection{Ambiguous Reactions: When ``Likes'' Fail to Signal Trust}
\label{section:4-2-4}
Participants expressed confusion about interpreting the meaning of reactions they receive to their posts, particularly when those reactions felt \inlinequote{obligatory} (\entry{19}). \entry{19} explained, \inlinequote{you feel more obligated to like a post\ldots [so] it just feels worse to post and not get as many likes.} Similarly, \codesign{08} remarked that when \inlinequote{people you're close with\ldots don't interact with your post,} they begin to question, \inlinequote{Is it really that boring that even people I know don't care?} While ``Likes'' may have been designed to signal validation, their meaning has become obscured in many platforms, making it an ineffective signal for reciprocity or trust.

To reduce ambiguity in social interactions, participants expressed interest in having a broader range of mechanisms for \textbf{intentional signaling} beyond effortless reactions such as ``Likes''. They cited examples from existing platforms where engagement from other users served as a clear, indisputable signal of care. For instance, the range and expressive nature of emoji reactions requires more deliberation than \inlinequote{just double tap[ing]}(\codesign{15}). When the chosen emoji aligns with the content, it sends clear signals that the actor actually viewed it. Comments were also valued as another form of clear signals of trust as they were seen as requiring individuals \inlinequote{go out of their way to be nice} (\codesign{08}). Participants also appreciated Instagram's Story Likes, describing them as \inlinequote{not obligatory at all} and as a feature that \inlinequote{shows that\ldots they decided to go out of their way to like it} (\entry{19}). To \codesign{13}, receiving Story Likes brings \inlinequote{a lot of trust[s]} given the private---and therefore more intimate, deliberate, and less likely performative---nature of those reactions. These reflections highlight that more nuanced, intentional reactions could serve as better social signals that reduce the ambiguity in boundary regulation.

\subsection{Low-Grace Culture: Barrier to Sustained Trust and Adaptive Boundary Setting}
\label{section:4-3}
In this section, we explore how social media cultivates a ``low-grace culture'' that discourages vulnerability or casual self-disclosure. This environment is characterized by an emphasis on polished self-presentation, judgments based on limited information, the risk of inadvertently burdening others by sharing, and an overall climate of distrust. These pressures inhibit the flexibility needed for adaptive boundary regulation, discouraging trust-building interactions.

\subsubsection{High-Stakes Presentation Expectations and Self-Censorship}
\label{section:4-3-1}
Participants frequently expressed fear about not measuring up to the presentation expectations of platforms such as Instagram. Especially on image-centric platforms, they felt compelled to share only highly polished moments. As \entry{18} stated, \inlinequote{Instagram is only for when the sun sets during the golden hour when I look my best.} This expectation discourages casual posting, as teens perceive that they \inlinequote{need to put significant effort into a post} (\entry{06}). Some worry that deviating from these norms might make them be judged as \inlinequote{underwhelming} or \inlinequote{monotonous} (\diary{11}), leaving them feeling vulnerable (\diary{13}).

To counteract these pressures, many participants expressed a preference for \textbf{low-stakes disclosure} (Figure \ref{fig:prototypes}(a)), which reduces the burden of excessive curation and supports more casual self-expression. They valued features that support ephemeral sharing, such as Instagram's Story or Snapchat posts, which feel \inlinequote{not as big of a deal} (\diary{16}) without \inlinequote{people saving or revisiting them} (\entry{06}). Participants often leveraged these features to share what they felt \inlinequote{doesn't really live up to the standards of what I post on Instagram} (\diary{07}). They also appreciated the support for casual, text-based updates on Bluesky and Twitter, or platforms that made interactions feel more like \inlinequote{having a conversation with someone} (\codesign{11}) rather than putting together a \inlinequote{portfolio} (\codesign{11}). Some participants suggested features like \inlinequote{a little status update} where people can \inlinequote{learn little facts about other people casually} (\codesign{05}).

These features help alleviate concerns about inadvertently exposing too much or misjudging boundaries by supporting and normalizing casual sharing rather than polished, high-effort posts. If casual sharing is the norm, then posting less curated content is less likely to be perceived as a disclosure of something deeply personal or intimate. Instead, it becomes part of a broader culture of mutual disclosure and relationship building, reducing the likelihood of boundary missteps and supporting better trust calibration among peers.

\subsubsection{Sparse Sharing and the Heightened Risk of Misrepresentation}
\label{section:4-3-2}
On platforms like Instagram, where participants perceived infrequent posting as the norm, limited content often becomes the primary basis for one's digital identity. This leaves users vulnerable to misrepresentation from acquaintances who may \inlinequote{make assumptions or misjudge [their] intentions} (\codesign{03}). With such sparse sharing, even a small number of posts can significantly influence how someone is perceived: \inlinequote{Somebody might pull up my Instagram and there are only, like, two posts and I look bad in both of them} (\codesign{12}). They also shared specific examples, such as posting extensively about a single topic, potentially coming across as being \inlinequote{obsessed} (\diary{12}) with that subject; sharing multiple photos from a memorable event might be interpreted as \inlinequote{bragging} (\entry{12}); or inquiring whether other students have completed their summer assignment could be misconstrued as attempting to \inlinequote{push} (\entry{12}) peers to do their work, despite that not being the intention of the post. Such concerns often led participants to lean toward self-censorship.

To mitigate the risk of trust erosion and promote more adaptive boundary regulation, participants suggested features to facilitate \textbf{contextual clarity} (Figure \ref{fig:prototypes} (d))---designs that provide additional context to reduce misunderstandings and support trust-building on platforms like Instagram. These included: indicating personal information such as interests on profiles (\diary{12}), emphasizing the importance of comprehensive information before making judgments (\codesign{03}), supporting explanatory captions for images beyond standard captions (\codesign{07}), and implementing a \inlinequote{social battery} indicator to help manage interaction expectations (\codesign{11}). By fostering mechanisms that encourage giving others the benefit of the doubt, such designs could prevent trust erosion and provide contexts for boundary regulation in online environments where full contextual understanding is not always possible.

\subsubsection{Nonconsensual Exposure and the Fear of Burdening Others}
\label{section:4-3-3}
On broadcast social media platforms, following a user implicitly consents to viewing their content, often leading to viewers feeling overwhelmed and \inlinequote{buried} (\codesign{12}) in posts. As \codesign{12} noted, \inlinequote{If I'm following 100 people and they're all sharing four times a day, then that's 400 things I have to click through.} Conversely, those that are sharing---aware of this dynamic from their own experiences---often fear \inlinequote{spamming} (\entry{03}, \codesign{13}) or \inlinequote{clogging} (\diary{16}) others' feeds. This fear of being perceived as burdensome or irritating was also evident in diary entries, where participants expressed hesitation \inlinequote{that people may think I'm oversharing or like posting too much} (\diary{11}), or regret after realizing, \inlinequote{[I] shared a TON of reels today on my close friends and public for some reason :crying-face:} (\diary{09}).

Participants sought ways to mitigate such unnecessary friction. Many desired \textbf{self-contained disclosure} (Figure \ref{fig:prototypes}(e)) that would allow them to share without imposing their content on others. These unobtrusive communication mechanisms included \inlinequote{a little status update} (\codesign{05}) or sharing minor personal interests on profile pages rather than in followers' feeds. One participant proposed a \inlinequote{red dot} (\codesign{17}) on user profiles to indicate new content subtly, allowing viewers to \inlinequote{not have to go look at it if they don't want to} (\codesign{17}). \codesign{05} saw parallels to Instagram's Story feature where \inlinequote{it doesn't notify people\ldots{} it's not like it goes on their feed\ldots{} it's just casual.} Participants perceived this as a way to foster connection without unnecessary friction and trust erosion by ensuring disclosure remains within a self-regulated and lower-stakes space.

\subsubsection{Absent Norms for Trust-Building and the Cycle of Distrust}
\label{section:4-3-4}
Participants observed social media spaces as often having a pervasive climate of distrust and the lack of measures to regulate such an environment. As \codesign{17} remarked, \inlinequote{On mainstream social media\ldots there's occasional positivity\ldots but mostly it feels draining}. 

They suggested that platforms could shift this culture by explicitly stating their platform expectations toward \textbf{trust-centered norms} (Figure \ref{fig:prototypes} (h)), such as explicitly stating that the platform is \inlinequote{a judgment-free zone} and communicating a clear message to users to \inlinequote{decrease judgment and be more kind} (\codesign{19}). They believed that such an approach would encourage users to \inlinequote{naturally be inclined to conform with the overall vibe and mission of the other users on the app} (\codesign{19}). Once a culture is established, a \inlinequote{selection bias} would occur, attracting like-minded individuals who align with these expectations (\codesign{19}). One participant expanded on this idea: 
\blockquote{I'm tired of how curated social media can be, but I still engage with it\ldots{} When the culture prioritizes authenticity and transparency, you'll get people like me, who are tired of curating, putting in the effort to be authentic and create this culture.}{\codesign{15}}
Similarly, another participant observed:
\blockquote{When a platform declares itself a judgment-free zone, it gives people the power to ensure both their own safety and the safety of others, creating a sense of control.}{\codesign{13}}
Participants emphasized that they believe that setting clear expectations, rules, and guidelines would foster meaningful changes toward safer environments. \codesign{17} noted that \inlinequote{teenagers my age listen when things are specifically told to them,} underscoring the importance of explicit communication in shaping platform norms.

Participants acknowledged that explicit guidelines could help mitigate negative interactions, but several also proposed additional features to actively \inlinequote{regulate toxic behaviors} as a necessary complement. They recognized that \inlinequote{there's always going to be someone that might try to ruin it} (\codesign{18}). To address this, \codesign{07} suggested a \inlinequote{negative comment filter} that users could toggle to block harmful content. Additionally, \codesign{19} proposed a reporting feature that would allow users to flag individuals \inlinequote{who may spread hurtful things or judge people.} While participants were mindful of the risk of implying the platform's distrust in its users, they acknowledged that tools for enforcing platform expectations could be valuable in strengthening their efforts to foster trust-enabling environments.
\section{Discussion}

\begin{figure*}[!ht]
    \centering
    \includegraphics[width=0.8\linewidth]{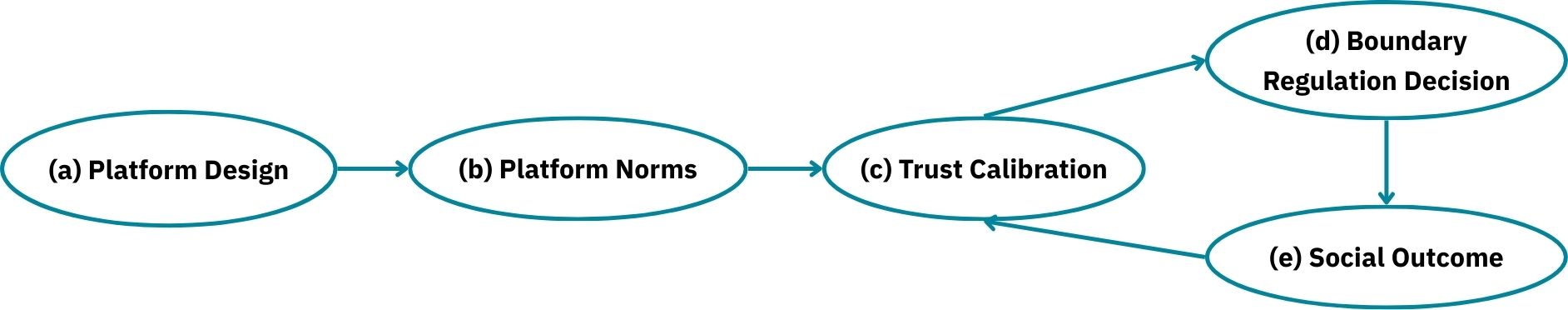}
    \caption{The proposed framework of Trust-Enabled Privacy.}
    \label{fig:framework}
\end{figure*}
\begin{table*}[!h]
\centering
\footnotesize
\caption{Details of Trust-Eroding and Trust-Building Processes in the Trust-Enabled Privacy Framework (Figure~\ref{fig:framework})}
\label{tab:framework-table}
\begin{tabular}{p{3cm} p{7cm} p{6.5cm}}
\toprule
\textbf{Component} & \textbf{Trust-Eroding} & \textbf{Trust-Building} \\
\midrule

\textbf{(a) Platform Design} & e.g.) effortless, obligatory reactions; non-consensual feed; image-forward sharing & Guided / Low-Stakes / Contextual / Self-Contained Disclosure; Contextual Clarity; Intentional Signaling; Mutual Commitment; Trust-Centered Norms \\
\midrule
\textbf{(b) Platform Norms} & 
\textbullet{} Communication Fog: ambiguous norms / loyalty / relevance / reactions \newline
\textbullet{} Low-Grace Culture: presentation expectations; social risk of misrepresentation; nonconsensual exposure; cycle of distrust & 
norms conducive to trust-building \\
\midrule
\textbf{(c) Trust Calibration} & lower trust & higher trust \\
\midrule
\textbf{(d) Boundary Regulation Decision} & 
withdrawal (i.e., non-disclosure); privacy resignation (i.e., oversharing); self-censored, infrequent sharing & 
casual, comfortable self-disclosure \\
\midrule
\textbf{(e) Social Outcome} & missed connections & relationship building and increased trust\\
\bottomrule
\end{tabular}
\end{table*}

Our study extends Waldman's theorization of privacy as a mechanism of trust~\cite{Waldman2018-dc} and demonstrates its empirical relevance in the social media context, especially with teens. We show how specific platform designs shape interpersonal trust, either reinforcing or undermining it, and how these trust dynamics unfold into concrete disclosure and boundary-regulation behaviors. By framing privacy as a shared, dynamic process rather than an individual burden, we propose design strategies that empower teens to regulate boundaries while fostering meaningful social connections. Specifically, our findings contribute (1) empirical insights into how teens navigate boundary regulation in social media environments based on trust levels, (2) identification of key design shortcomings that undermine trust-building, and (3) a set of design interventions---including both adaptations of existing mechanisms and novel ideas---that reflect teen users' perspectives on what is most effective for supporting boundary regulation and relational needs.

Importantly, some of the design ideas and prototypes we discuss are inspired by existing features or implementations. However, our work extends their relevance by demonstrating how they contribute to \textit{\textbf{trust-enabled privacy}}. Rather than viewing these features in isolation, we show how they function to foster relationship-building, reduce social risk, and support dynamic boundary regulation. By centering \textbf{\textit{trust}} as the mechanism that integrates boundary regulation and relationship-building, our work offers a framework for designing social media platforms that are developmentally and socially attuned.

\subsection{Developmentally Sensitive Privacy Design: Relationship Building and Empowerment}
Prior work has emphasized that teens' privacy concerns are deeply relational~\cite{zhao2022understanding} and that boundary regulation is a co-managed, ongoing process in networked environments~\cite{marwick2014networked}. The ability to control what they share with specific audiences---ensuring that disclosures reach trusted friends while remaining shielded from unintended viewers such as teachers or potential employers---is essential~\cite{Sleeper-2013-PostWasntFacebook-m, Stutzman-2012-BoundaryRegulationMedia-n}. However, privacy is not merely about separating trusted and untrusted audiences; it is a fluid, iterative process shaped by ongoing interactions, trust-building, and trust erosion over time. Framing privacy solely as information restriction---an approach often reinforced by prevention-focused online safety strategies~\cite{Wisniewski-2018-PrivacyParadox-l}---can push teens toward withdrawal and disengagement~\cite{kim2025privacysocialnormsystematically}, ultimately denying them opportunities to develop and negotiate social relationships online. Given the developmental significance of relationship-building during adolescence~\cite{Davis2012-bq}, privacy protections should not come at the cost of self-disclosure, as sharing is a fundamental mechanism for forming and strengthening social connections. Worse, rigid privacy models that force teens into a false trade-off between privacy and connection may leave them prioritizing connection at the expense of privacy rather than enabling them to manage boundaries in a way that adapts to shifting relationships and evolving trust.

Teens, highly attuned to peer norms and the nuances of social dynamics~\cite{Somerville2013-zf, Steinberg2005-qy}, often perceive online self-disclosure as high-stakes~\cite{Yau2019-ab}. While increased sharing might seem risky, self-disclosure and privacy are not inherently at odds. Rather, privacy functions as a foundation that enables confident sharing, shaping the conditions under which disclosure feels safe or precarious. In environments of distrust, teens share less yet feel more vulnerable when privacy violations occur~\cite{Petronio2002-ce, Williams2023-jm}. Conversely, when privacy norms are well-supported and trust is reinforced, teens not only feel more comfortable disclosing but are also more likely to respect others' privacy and assert their own boundaries~\cite{kim2025privacysocialnormsystematically, Waldman2018-dc, GhaiumyAnaraky-2021-DiscloseNotAdults-l}. By fostering an environment where privacy is actively negotiated rather than rigidly enforced, platforms can encourage a culture of mutual respect---helping teens navigate boundary regulation with greater agency and confidence.

\subsection{Trust-Enabled Privacy Framework}
Figure~\ref{fig:framework} illustrates how platform design initiates a cascading process that shapes boundary regulation through trust. Design decisions establish platform norms, such as acceptable levels of sharing, expected reactions, and perceived audience boundaries, which influence how teens interpret their social environment and assess audience trustworthiness. This trust calibration then informs decisions to share, withdraw, or self-censor, leading to social outcomes that feed back into their sense of trust. The framework contrasts trust-eroding pathways, fueled by features like non-consensual feeds and obligatory reactions that create ambiguous norms, with trust-building designs, such as contextual audience segmentation and mutual opt-in spaces, which promote safer disclosure and reinforce relational trust.

Taken together, Figure~\ref{fig:framework} and Table~\ref{tab:framework-table} show that trust-enabled boundary regulation is not just a matter of individual behavior but an ecosystem-level outcome that starts and often breaks down with platform design. Many of the current social media designs implicitly assume or actively cultivate a low-trust environment where privacy must be actively defended rather than mutually respected. By centering trust within the privacy process, this framework offers a blueprint for future interventions that support adaptive, socially aware boundary regulation. Instead of treating privacy as a binary on/off switch, trust-enabled privacy recognizes it as a developmental, relational process, one that teens need to be empowered to actively navigate and that platforms must meaningfully support.

\subsection{Design Guidelines for Teen User Social Media Privacy}
Our study shows the need for design interventions that recognize privacy as a relational process rather than individual responsibility. We propose four key design guidelines to support teens in managing privacy through trust-based interactions; details and example prototypes illustrating these guidelines can be found in Appendix \ref{appendix-A}.

\parHeading{Contextualized and Bilateral Agreement in Sharing} Platforms should support more intentional and segmented sharing by letting viewers opt into specific content, rather than receiving posts passively. This model helps both sharers and audiences set clearer mutual expectations around visibility and engagement, easing pressure on sharers and reducing ambiguity. Interest-based, self-curated spaces can also encourage teens to share more thoughtfully within relevant circles. Giving viewers a role in managing boundaries would help align privacy settings with evolving relationships.

\parHeading{Options for Thoughtful, Low-Stakes Sharing} When contextual sharing is hard, such as with large, vague audiences, platforms should offer low-pressure ways to disclose personal updates. Features like ephemeral posts, subtle status updates (e.g., profile music), and casual sharing norms let users open up without feeling overly exposed. To maintain trust, platforms should minimize friction, such as preventing posts from auto-appearing in feeds or allowing users to add context ahead of time.

\parHeading{Better Social Signals for Trust-Building and Boundary Regulation} Trust builds through repeated, meaningful interaction, but current designs often lack good signals for it. Platforms should offer more contextual feedback, beyond ``Likes,'' such as tailored reactions, quiet acknowledgments, or small affirmations that reassure sharers. Soft cues about who's viewing and engaging (e.g., shared interests or signs of reciprocity) can also help users feel more confident in setting boundaries.

\parHeading{Towards Collective Norms for Trust-Building} Privacy is fundamentally norm-based. Rather than placing the full burden of privacy management on individual users, platforms should establish shared norms that define and reinforce responsible boundary regulation. By promoting collective norms---through platform-wide standards, accountability tools, and proactive enforcement---social media can better support users in managing privacy within complex social dynamics.

\subsection{Limitations and Future Work}
We recruited participants through Instagram ads, so all of them were Instagram users. Although the study did not target any one platform, most examples participants gave centered on Instagram. This might suggest that teens see Instagram as the most representative of social media, though our recruitment likely influenced this since frequent Instagram users were more likely to participate. All participants were based in the U.S., which limits how broadly our findings apply, especially given that social media use and social norms vary across cultures. Our sample also skewed toward girls and older teens.

Our research centered on facilitating meaningful self-disclosure among teens, particularly with peers they do not yet fully trust. However, it is important to note that social desirability bias~\cite{fisher1993social} may have affected participants' responses. The issue of ``oversharing'' to the point of compromising privacy without receiving social validation remains prevalent. We did not explore privacy concerns like screenshots, privacy breaches, or the broader risks of content being shared out of context---issues that remain critical.



Future studies should develop and empirically test the design concepts introduced in this paper and investigate the specifics of what norms conducive to trust-building (Table~\ref{tab:framework-table}) entail.
\section{Conclusion}
This study highlights the critical role of trust-enabled privacy in shaping adolescent social media experiences. Our findings illustrate how communication fog and low-grace culture hinder self-disclosure, leading teens to default to disengagement rather than risk uncertainty. In response, we propose design interventions that empower teens to regulate their boundaries dynamically, ensuring privacy is not merely about restricting access but about facilitating gradual trust-building in evolving social contexts. Rather than treating privacy as an individual responsibility, platforms should reimagine boundary regulation as a shared, social process. Teens advocate for contextualized audience control, mechanisms for mutual trust reinforcement, and lightweight, low-risk sharing options that align with their relational needs. By integrating these insights, platforms can move beyond rigid information-control models and instead cultivate social media environments that actively support trust development.

\section*{Acknowledgments}
The authors would like to acknowledge the CERES network, the University of Washington Global Innovation Funds (GIF), and the University of Washington Student Technology Funds (STF), which provided support for this work. We additionally thank the anonymous reviewers for their detailed feedback and the participants for sharing their thoughts. We truly appreciate all the help. Alexis Hiniker is a special government employee for the Federal Trade Commission. The content expressed in this manuscript does not reflect the views of the Commission or any of the Commissioners.

\bibliographystyle{plain}
\bibliography{references}

\begin{thebibliography}{10}

\bibitem{google-docs}
Google docs: Online document editor.
\newblock \url{https://www.google.com/docs/about/}.
\newblock Accessed: 2024-9-12.

\bibitem{Other-Other-NYState2023-S7694A-p}
{NY} state senate bill 2023-{S7694A}.
\newblock \url{https://www.nysenate.gov/legislation/bills/2023/S7694/amendment/A}.
\newblock Accessed: 2025-1-20.

\bibitem{Other-Other-Osf-n}
{OSF}.
\newblock \url{https://osf.io/dashboard}.
\newblock Accessed: 2025-5-22.

\bibitem{miro}
The visual collaboration platform for every team.
\newblock \url{https://miro.com/}.
\newblock Accessed: 2023-7-18.

\bibitem{atlas}
{ATLAS.ti}.
\newblock \url{https://atlasti.com/}, January 2024.
\newblock Accessed: 2024-1-15.

\bibitem{Abaquita-2020-PrivacyNormsIntegrity-m}
Denielle Abaquita, Paritosh Bahirat, Karla~A Badillo-Urquiola, and Pamela Wisniewski.
\newblock Privacy norms within the internet of things using contextual integrity.
\newblock In {\em Companion of the 2020 ACM International Conference on Supporting Group Work}, New York, NY, USA, January 2020. ACM.

\bibitem{Acquisti2017-ez}
Alessandro Acquisti, Idris Adjerid, Rebecca Balebako, Laura Brandimarte, Lorrie~Faith Cranor, Saranga Komanduri, Pedro~Giovanni Leon, Norman Sadeh, Florian Schaub, Manya Sleeper, Yang Wang, and Shomir Wilson.
\newblock Nudges for privacy and security: Understanding and assisting users' choices online.
\newblock {\em ACM Comput. Surv.}, 50(3):1--41, August 2017.

\bibitem{Agha2023-dv}
Zainab Agha.
\newblock To nudge or not to nudge: {Co-Designing} and evaluating the effectiveness of adolescent online safety nudges.
\newblock In {\em Proceedings of the 22nd Annual {ACM} Interaction Design and Children Conference}, Idc '23, pages 760--763, New York, NY, USA, June 2023. Association for Computing Machinery.

\bibitem{Agha2023-mu}
Zainab Agha, Karla Badillo-Urquiola, and Pamela~J Wisniewski.
\newblock ``strike at the root'': Co-designing {Real-Time} social media interventions for adolescent online risk prevention.
\newblock {\em Proc. ACM Hum.-Comput. Interact.}, 7(Cscw1):1--32, April 2023.

\bibitem{Akter-2024-ExaminingCaregivingSecurity-n}
Mamtaj Akter, Jess Kropczynski, H~Lipford, and Pamela~J Wisniewski.
\newblock Examining caregiving roles to differentiate the effects of using a mobile app for community oversight for privacy and security.
\newblock {\em ArXiv}, abs/2409.02364, September 2024.

\bibitem{Akter-2023-EvaluatingImpactSecurity-h}
Mamtaj Akter, Madiha Tabassum, Md~Nazmus~Sakib Miazi, Leen~A Alghamdi, Jessica Kropczynski, P~Wisniewski, and H~Lipford.
\newblock Evaluating the impact of community oversight for managing mobile privacy and security.
\newblock {\em Symp Usable Priv Secur}, abs/2306.02289:437--456, June 2023.

\bibitem{altman1973social}
Irwin Altman and Dalmas~A Taylor.
\newblock {\em Social penetration: The development of interpersonal relationships.}
\newblock Holt, Rinehart \& Winston, 1973.

\bibitem{Anderson2023-qk}
Monica Anderson.
\newblock Teens, social media and technology 2023.
\newblock \url{https://www.pewresearch.org/internet/2023/12/11/teens-social-media-and-technology-2023/}, December 2023.
\newblock Accessed: 2024-9-3.

\bibitem{Armsden1987-ai}
G~C Armsden and M~T Greenberg.
\newblock The inventory of parent and peer attachment: Individual differences and their relationship to psychological well-being in adolescence.
\newblock {\em J. Youth Adolesc.}, 16(5):427--454, October 1987.

\bibitem{Ayalon-2013-RetrospectivePrivacyNetworks-t}
Oshrat Ayalon and Eran Toch.
\newblock Retrospective privacy: managing longitudinal privacy in online social networks.
\newblock In {\em Proceedings of the Ninth Symposium on Usable Privacy and Security}, New York, NY, USA, July 2013. ACM.

\bibitem{Barkhuus-2012-MismeasurementPrivacyHCI-d}
Louise Barkhuus.
\newblock The mismeasurement of privacy: using contextual integrity to reconsider privacy in {HCI}.
\newblock In {\em Proceedings of the SIGCHI Conference on Human Factors in Computing Systems}, New York, NY, USA, May 2012. ACM.

\bibitem{BoydDanah2014ICTS}
danah boyd.
\newblock {\em It's Complicated: The Social Lives of Networked Teens}.
\newblock Yale University Press, New York, 1 edition, 2014.

\bibitem{braun2019reflecting}
Virginia Braun and Victoria Clarke.
\newblock Reflecting on reflexive thematic analysis.
\newblock {\em Qualitative research in sport, exercise and health}, 11(4):589--597, 2019.

\bibitem{burgstahler2010universal}
Sheryl~E Burgstahler and Rebecca~C Cory.
\newblock {\em Universal design in higher education: From principles to practice}.
\newblock Harvard Education Press, 2010.

\bibitem{Cho-2018-CollectivePrivacyValidation-h}
Hichang Cho, Bart Knijnenburg, Alfred Kobsa, and Yao Li.
\newblock Collective privacy management in social media: A cross-cultural validation.
\newblock {\em ACM Trans. Comput. Hum. Interact.}, 25(3):1--33, June 2018.

\bibitem{Choksi-2024-PrivacyGroupsMatters-b}
Madiha~Zahrah Choksi, Ero Balso, Frauke Kreuter, and Helen Nissenbaum.
\newblock Privacy for groups online: Context matters.
\newblock {\em Proc. ACM Hum. Comput. Interact.}, 8(CSCW2):1--23, November 2024.

\bibitem{Chouhan-2019-Co-designingCommunityTogether-x}
Chhaya Chouhan, Christy~M LaPerriere, Zaina Aljallad, Jess Kropczynski, Heather Lipford, and Pamela~J Wisniewski.
\newblock Co-designing for community oversight: Helping people make privacy and security decisions together.
\newblock {\em Proc. ACM Hum. Comput. Interact.}, 3(CSCW):1--31, November 2019.

\bibitem{Christofides-2012-HeyMomAdults-o}
Emily Christofides, Amy Muise, and Serge Desmarais.
\newblock Hey mom, what’s on your facebook? comparing facebook disclosure and privacy in adolescents and adults.
\newblock {\em Soc. Psychol. Personal. Sci.}, 3(1):48--54, January 2012.

\bibitem{Colnago-2023-ThereReverseBehaviors-j}
Jessica Colnago, Lorrie Cranor, and Alessandro Acquisti.
\newblock Is there a reverse privacy paradox? an exploratory analysis of gaps between privacy perspectives and privacy-seeking behaviors.
\newblock {\em Proc. Priv. Enhancing Technol.}, 2023(1):455--476, January 2023.

\bibitem{Davis2012-bq}
Katie Davis.
\newblock Friendship 2.0: adolescents' experiences of belonging and self-disclosure online.
\newblock {\em J. Adolesc.}, 35(6):1527--1536, December 2012.

\bibitem{de2020contextualizing}
Ralf De~Wolf.
\newblock Contextualizing how teens manage personal and interpersonal privacy on social media.
\newblock {\em New media \& society}, 22(6):1058--1075, 2020.

\bibitem{DeWolf-2014-ManagingPrivacyFacebook-c}
Ralf De~Wolf, Koen Willaert, and Jo~Pierson.
\newblock Managing privacy boundaries together: Exploring individual and group privacy management strategies in facebook.
\newblock {\em Comput. Human Behav.}, 35:444--454, June 2014.

\bibitem{derlega1979self}
Valerian~J Derlega, Janusz Grzelak, and GJ~Chelune.
\newblock Self-disclosure: Origins, patterns, and implications of openness in interpersonal relationships.
\newblock {\em Appropriateness of self-disclosure}, pages 151--176, 1979.

\bibitem{derlega1993self}
Valerian~J Derlega, Sandra Metts, Sandra Petronio, and Stephen~T Margulis.
\newblock {\em Self-disclosure.}
\newblock Sage Publications, Inc, 1993.

\bibitem{Dienlin-2015-PrivacyParadoxBehaviors-w}
Tobias Dienlin and Sabine Trepte.
\newblock Is the privacy paradox a relic of the past? an in-depth analysis of privacy attitudes and privacy behaviors: The relation between privacy attitudes and privacy behaviors.
\newblock {\em Eur. J. Soc. Psychol.}, 45(3):285--297, April 2015.

\bibitem{Ellison2011-ld}
Nicole~B Ellison, Charles Steinfield, and Cliff Lampe.
\newblock Connection strategies: Social capital implications of facebook-enabled communication practices.
\newblock {\em New Media Soc.}, 13(6):873--892, September 2011.

\bibitem{fisher1993social}
Robert~J Fisher.
\newblock Social desirability bias and the validity of indirect questioning.
\newblock {\em Journal of consumer research}, 20(2):303--315, 1993.

\bibitem{GhaiumyAnaraky-2021-DiscloseNotAdults-l}
Reza Ghaiumy~Anaraky, Kaileigh~Angela Byrne, Pamela~J Wisniewski, Xinru Page, and Bart Knijnenburg.
\newblock To disclose or not to disclose: Examining the privacy decision-making processes of older vs. younger adults.
\newblock In {\em Proceedings of the 2021 CHI Conference on Human Factors in Computing Systems}, New York, NY, USA, May 2021. ACM.

\bibitem{hargittai2016can}
Eszter Hargittai and Alice Marwick.
\newblock ``what can i really do?'' explaining the privacy paradox with online apathy.
\newblock {\em International journal of communication}, 10:21, 2016.

\bibitem{Hogan2010-oh}
Bernie Hogan.
\newblock The presentation of self in the age of social media: Distinguishing performances and exhibitions online.
\newblock {\em Bull. Sci. Technol. Soc.}, 30(6):377--386, 2010.

\bibitem{Humbert-2019-SurveyInterdependentPrivacy-f}
Mathias Humbert, Benjamin Trubert, and Kévin Huguenin.
\newblock A survey on interdependent privacy.
\newblock {\em ACM Comput. Surv.}, 52(6):1--40, October 2019.

\bibitem{Joinson2010-cf}
Adam~N Joinson, Ulf-Dietrich Reips, Tom Buchanan, and Carina B~Paine Schofield.
\newblock Privacy, trust, and {Self-Disclosure} online.
\newblock {\em Human--Computer Interaction}, 25(1):1--24, February 2010.

\bibitem{kim2025socialmediafeellike}
JaeWon Kim, Hyunsung Cho, Fannie Liu, and Alexis Hiniker.
\newblock Social media should feel like minecraft, not instagram: 3d gamer youth visions for meaningful social connections through fictional inquiry, 2025.

\bibitem{kim2025privacysocialnormsystematically}
JaeWon Kim, Soobin Cho, Robert Wolfe, Jishnu~Hari Nair, and Alexis Hiniker.
\newblock Privacy as social norm: Systematically reducing dysfunctional privacy concerns on social media, 2025.

\bibitem{10.1145/3686909}
JaeWon Kim, Robert Wolfe, Ishita Chordia, Katie Davis, and Alexis Hiniker.
\newblock "sharing, not showing off": How bereal approaches authentic self-presentation on social media through its design.
\newblock {\em Proc. ACM Hum.-Comput. Interact.}, 8(CSCW2), November 2024.

\bibitem{Kim-2024-AustraliaBarred-y}
Victoria Kim.
\newblock Australia has barred everyone under 16 from social media. will it work?
\newblock {\em The New York Times}, November 2024.

\bibitem{knijnenburg2022modern}
Bart~P Knijnenburg, Xinru Page, Pamela Wisniewski, Heather~Richter Lipford, Nicholas Proferes, and Jennifer Romano.
\newblock {\em Modern socio-technical perspectives on privacy}.
\newblock Springer Nature, 2022.

\bibitem{Lam2012-ct}
S~K Lam and J~Riedl.
\newblock Are our online ``friends'' really friends?
\newblock {\em Computer}, 45(1):91--93, January 2012.

\bibitem{Lenhart2015-cm}
Amanda Lenhart.
\newblock Chapter 4: Social media and friendships.
\newblock \url{https://www.pewresearch.org/internet/2015/08/06/chapter-4-social-media-and-friendships/}, August 2015.
\newblock Accessed: 2024-4-28.

\bibitem{meta-sue}
Cristiano Lima-Strong and Naomi Nix.
\newblock 41 states sue meta, claiming instagram, facebook are addictive, harm kids.
\newblock {\em The Washington Post}, October 2023.

\bibitem{Lowens-2025-MisalignmentsDemographicFacebook-d}
Byron Lowens, Sean Scarnecchia, Jane Im, Tanisha Afnan, Annie Chen, Yixin Zou, and Florian Schaub.
\newblock Misalignments and demographic differences in expected and actual privacy settings on facebook.
\newblock {\em Proc. Priv. Enhancing Technol.}, 2025(1):456--471, January 2025.

\bibitem{Malhotra-2024-"whatPostPlaces-u}
Pranav Malhotra.
\newblock “what you post in the group stays in the group”: Examining the affordances of bounded social media places.
\newblock {\em Soc. Media Soc.}, 10(3), July 2024.

\bibitem{Malkin-2022-RuntimePermissionsAssistants-p}
Nathan Malkin, David Wagner, and Serge Egelman.
\newblock Runtime permissions for privacy in proactive intelligent assistants.
\newblock In {\em Eighteenth Symposium on Usable Privacy and Security (SOUPS 2022)}, pages 633--651, 2022.

\bibitem{marwick2014networked}
Alice~E Marwick and Danah Boyd.
\newblock Networked privacy: How teenagers negotiate context in social media.
\newblock {\em New media \& society}, 16(7):1051--1067, 2014.

\bibitem{mcpherson2006social}
Miller McPherson, Lynn Smith-Lovin, and Matthew~E Brashears.
\newblock Social isolation in america: Changes in core discussion networks over two decades.
\newblock {\em American sociological review}, 71(3):353--375, 2006.

\bibitem{Mondal-2019-MovingBeyondMedia-p}
Mainack Mondal, Günce~Su Yilmaz, Noah Hirsch, Mohammad~Taha Khan, Michael Tang, Christopher Tran, Chris Kanich, Blase Ur, and Elena Zheleva.
\newblock Moving beyond set-it-and-forget-it privacy settings on social media.
\newblock In {\em Proceedings of the 2019 ACM SIGSAC Conference on Computer and Communications Security}, New York, NY, USA, November 2019. ACM.

\bibitem{Namara-2022-EffectivenessAdaptationSites-y}
Moses Namara, Henry Sloan, and Bart~P Knijnenburg.
\newblock The effectiveness of adaptation methods in improving user engagement and privacy protection on social network sites.
\newblock {\em Proc. Priv. Enhancing Technol.}, 2022(1):629--648, January 2022.

\bibitem{Nguyen2012-cc}
Melanie Nguyen, Yu~Sun Bin, and Andrew Campbell.
\newblock Comparing online and offline {Self-Disclosure}: A systematic review.
\newblock {\em Cyberpsychol. Behav. Soc. Netw.}, 15(2):103--111, 2012.

\bibitem{Nguyen2009-ri}
Viet-An Nguyen, Ee-Peng Lim, Jing Jiang, and Aixin Sun.
\newblock To trust or not to trust? predicting online trusts using trust antecedent framework.
\newblock In {\em 2009 Ninth IEEE International Conference on Data Mining}, pages 896--901. IEEE, December 2009.

\bibitem{Nissenbaum-2004-PrivacyContextualIntegrity-j}
H~Nissenbaum.
\newblock Privacy as contextual integrity.
\newblock {\em Washington Law Review}, 79:119--157, February 2004.

\bibitem{Paci-2019-SurveyAccessSystems-a}
Federica Paci, Anna Squicciarini, and Nicola Zannone.
\newblock Survey on access control for community-centered collaborative systems.
\newblock {\em ACM Comput. Surv.}, 51(1):1--38, January 2019.

\bibitem{Page-2019-PragmaticToolFeatures-p}
Xinru Page, Reza Ghaiumy~Anaraky, Bart~P Knijnenburg, and Pamela~J Wisniewski.
\newblock Pragmatic tool vs. relational hindrance: Exploring why some social media users avoid privacy features.
\newblock {\em Proc. ACM Hum. Comput. Interact.}, 3(CSCW):1--23, November 2019.

\bibitem{Petronio2002-ce}
Sandra Petronio.
\newblock {\em Boundaries of privacy: Dialectics of disclosure}.
\newblock SUNY series in Communication Studies. State University of New York Press, Albany, NY, October 2002.

\bibitem{Prinstein2023-pd}
Mitch Prinstein.
\newblock Written testimony.
\newblock Technical report, American Psychological Association (APA), February 2023.

\bibitem{Putnam2000-ic}
R~Putnam.
\newblock Bowling alone: the collapse and revival of american community.
\newblock {\em Conf Comput Support Cooperative Work}, page 357, 2000.

\bibitem{Reddy2024-ic}
Ananya Reddy and Priya~C Kumar.
\newblock `a teaspoon of authenticity': Exploring how young adults {BeReal} on social media.
\newblock In {\em Proceedings of the {CHI} Conference on Human Factors in Computing Systems}, number Article 907 in CHI '24, pages 1--14, New York, NY, USA, May 2024. Association for Computing Machinery.

\bibitem{saunders2018saturation}
Benjamin Saunders, Julius Sim, Tom Kingstone, Shula Baker, Jackie Waterfield, Bernadette Bartlam, Heather Burroughs, and Clare Jinks.
\newblock Saturation in qualitative research: exploring its conceptualization and operationalization.
\newblock {\em Quality \& quantity}, 52:1893--1907, 2018.

\bibitem{Schnitzler-2020-SoKManagingData-h}
Theodor Schnitzler, Muhammad~Shujaat Mirza, Markus Dürmuth, and Christina Pöpper.
\newblock {SoK}: Managing longitudinal privacy of publicly shared personal online data.
\newblock {\em Proc. Priv. Enhancing Technol.}, 2021:229--249, November 2020.

\bibitem{Sleeper-2013-PostWasntFacebook-m}
Manya Sleeper, Rebecca Balebako, Sauvik Das, Amber~Lynn McConahy, Jason Wiese, and Lorrie~Faith Cranor.
\newblock The post that wasn't: exploring self-censorship on facebook.
\newblock In {\em Proceedings of the 2013 conference on Computer supported cooperative work}, CSCW '13, pages 793--802, New York, NY, USA, February 2013. Association for Computing Machinery.

\bibitem{Somerville2013-zf}
Leah~H Somerville.
\newblock Special issue on the teenage brain: Sensitivity to social evaluation.
\newblock {\em Curr. Dir. Psychol. Sci.}, 22(2):121--127, April 2013.

\bibitem{Steinberg2005-qy}
Laurence Steinberg.
\newblock Cognitive and affective development in adolescence.
\newblock {\em Trends Cogn. Sci.}, 9(2):69--74, February 2005.

\bibitem{Stutzman-2012-BoundaryRegulationMedia-n}
Frederic Stutzman and Woodrow Hartzog.
\newblock Boundary regulation in social media.
\newblock In {\em Proceedings of the ACM 2012 conference on Computer Supported Cooperative Work}, CSCW '12, pages 769--778, New York, NY, USA, February 2012. Association for Computing Machinery.

\bibitem{Waldman2016-em}
A~Waldman.
\newblock Privacy, sharing, and trust: The facebook study.
\newblock {\em Case West. Reserve Law Rev.}, 67:193, June 2016.

\bibitem{Waldman2018-dc}
Ari~Ezra Waldman.
\newblock {\em Privacy as Trust: Information Privacy for an Information Age}.
\newblock Cambridge University Press, March 2018.

\bibitem{walther2011introduction}
Joseph~B Walther.
\newblock Introduction to privacy online.
\newblock In {\em Privacy online: Perspectives on privacy and self-disclosure in the social web}, pages 3--8. Springer, 2011.

\bibitem{Weinstein2022-rh}
Emily Weinstein and Carrie James.
\newblock {\em Behind Their Screens: What Teens Are Facing (and Adults Are Missing)}.
\newblock MIT Press, August 2022.

\bibitem{Williams2023-jm}
Olivia~A Williams, Yee-Yin Choong, and Kerrianne Buchanan.
\newblock Youth understandings of online privacy and security: A dyadic study of children and their parents.
\newblock {\em Symp Usable Priv Secur}, pages 399--416, 2023.

\bibitem{Wisniewski-2018-PrivacyParadox-l}
Pamela Wisniewski.
\newblock The privacy paradox of adolescent online safety: A matter of risk prevention or risk resilience?
\newblock {\em IEEE Secur. Priv.}, 16(2):86--90, 2018.

\bibitem{Wisniewski-2015-GiveSocialWant-z}
Pamela Wisniewski, A~K M~Najmul Islam, Bart~P Knijnenburg, and Sameer Patil.
\newblock Give social network users the privacy they want.
\newblock In {\em Proceedings of the 18th ACM Conference on Computer Supported Cooperative Work \& Social Computing}, New York, NY, USA, February 2015. ACM.

\bibitem{Wisniewski-2016-FramingMeasuringUsers-w}
Pamela Wisniewski, A~K Najmul~Islam, Heather~Richter Lipford, and David~C Wilson.
\newblock Framing and measuring multi-dimensional interpersonal privacy preferences of social networking site users.
\newblock {\em Communications of the Association for Information Systems}, 38(1):10, 2016.

\bibitem{Yau2019-ab}
Joanna~C Yau and Stephanie~M Reich.
\newblock ``it's just a lot of work'': Adolescents' {Self-Presentation} norms and practices on facebook and instagram.
\newblock {\em J. Res. Adolesc.}, 29(1):196--209, 2019.

\bibitem{zhao2022understanding}
Dorothy Zhao, Mikako Inaba, and Andr{\'e}s Monroy-Hern{\'a}ndez.
\newblock Understanding teenage perceptions and configurations of privacy on instagram.
\newblock {\em Proceedings of the ACM on Human-Computer Interaction}, 6(Cscw2):1--28, 2022.

\end{thebibliography}

\appendix

\onecolumn
\section{Appendix: Entry Interview Miro Board Exercise}
\label{ref:miro}

\begin{figure*}[!h]
    \centering
    \includegraphics[width=1.0\linewidth]{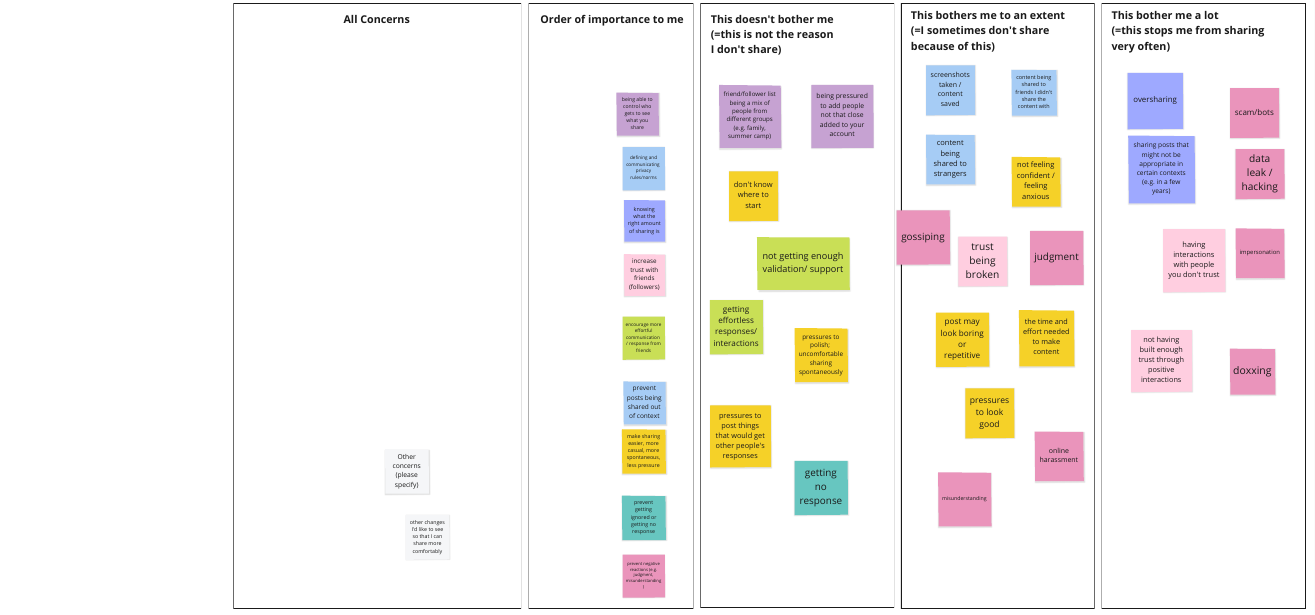}
    \caption{Screenshot of P01's Miro Boards Used During Co-Design (Exit) Interviews.}
\end{figure*}

\section{Appendix: Design Ideas and Prototypes}
\label{appendix-A}
\label{fig:prototypes}

\begin{figure*}[!h]
    \centering
    \includegraphics[width=0.8\linewidth]{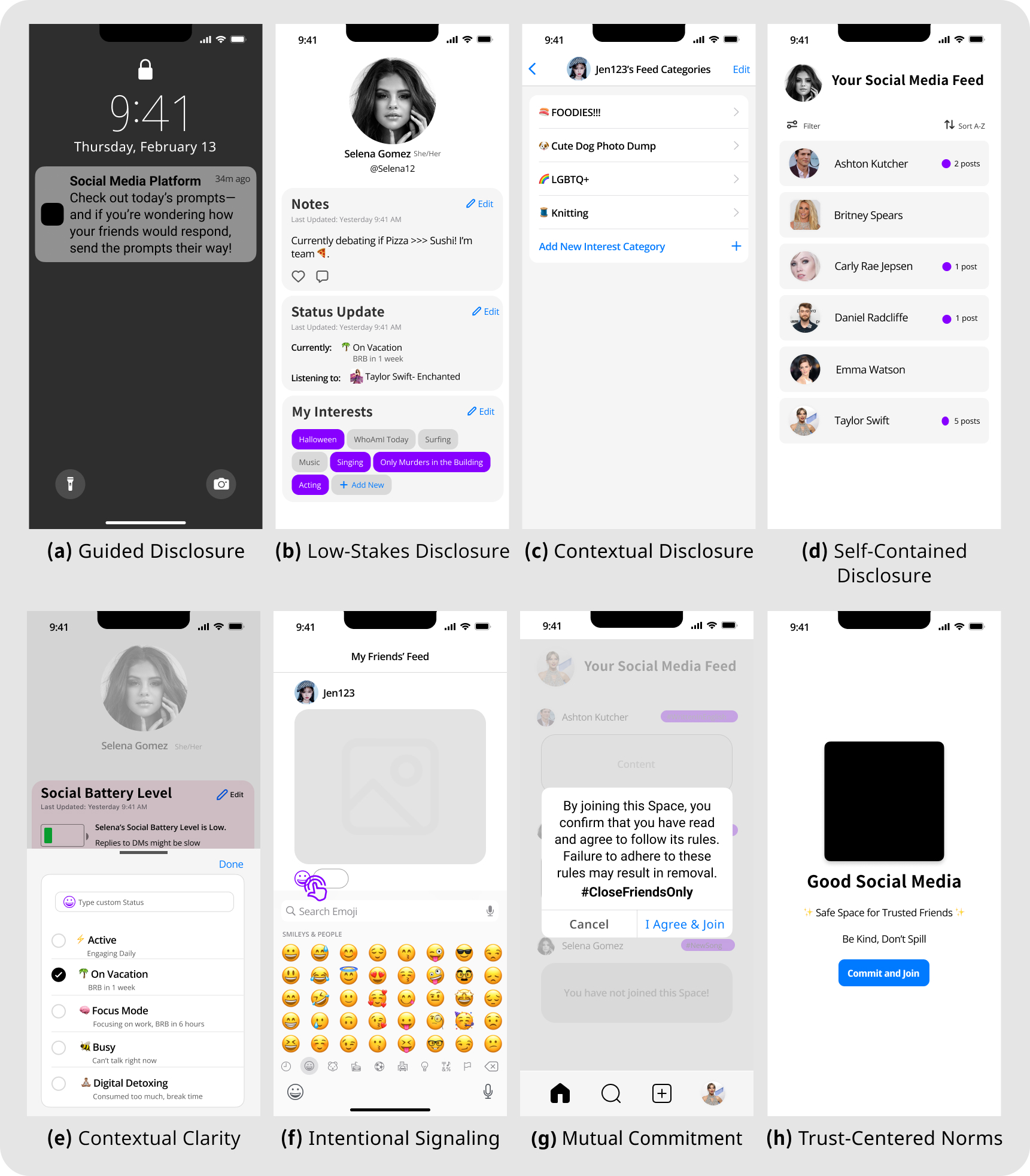}
    \caption{High-fidelity mock-ups of the eight design approaches derived from the co-design study with teen participants. A higher-level taxonomy of the designs is available in Table \ref{tab:designs}. (a) Provides structured prompts that reduce ambiguity around what's socially acceptable to post, helping users navigate implicit norms; (b) Fosters low-pressure sharing through casual status updates (e.g., ``pizza >>> sushi'') and structured ones (e.g., with multiple choice options), reducing the burden of curating polished content and supporting low-stakes disclosure; (c) Enables segmentation of posts into interest-based categories like ``Cute Dog Photo Dump'' or ``LGBTQ+,'' ensuring content is directed to the right audience and reducing ambiguity about relevance and who sees what; (d) Offers each user a dedicated sharing space from the outset, so they can post freely without fear of ``clogging'' others' feeds; (e) Allows users to communicate their availability or current mode (e.g., ``On Vacation,'' ``Digital Detoxing'') with contextual status indicators, allowing users to preemptively explain their communication boundaries, minimizing misunderstandings and social misjudgments; (f) Encourages more expressive and deliberate engagement by offering a wide range of emoji reactions, moving beyond generic `likes' to support nuanced, deliberate, and context-specific signaling; (g) Enforces mutual accountability by requiring users to agree to a space's rules before joining, reinforcing shared responsibility between sharers and viewers; (h) Cultivates a trusted environment by requiring users to agree to clear community guidelines before joining, enabling aligned participation and reinforcing shared norms.}
\end{figure*}

\begin{table*}[!h]
\centering
\small
\caption{Taxonomy of designs derived from the co-design study.}
\label{tab:designs}
\begin{tabular}{p{3.5cm} p{10.5cm} p{1.5cm}}
\toprule
\textbf{Overall Design Idea} & \textbf{Design Solution and Corresponding Trust-Eroding Platform Norm Addressed} & \textbf{Example Prototype} \\ 
\midrule

\textbf{Guided Disclosure} & Helps clarify \textbf{\textit{Ambiguous Norms}} [Section \ref{section:4-2-1}] by providing clear guidelines and expectations around posting, reducing uncertainty and the need for users to interpret implicit social expectations independently & Figure \ref{fig:prototypes}(a) \\

\textbf{Mutual Commitment} & Addresses \textbf{\textit{Ambiguous Loyalty}} [Section \ref{section:4-2-2}] by allowing viewers to opt in or out of content while requiring them to agree to the space's rules before joining, ensuring mutual commitment and accountability for both the sharer and the audience. & Figure \ref{fig:prototypes}(g) \\

\textbf{Contextual Disclosure} & Counteracts \textbf{\textit{Ambiguous Relevance}} [Section \ref{section:4-2-3}] by enabling segmentation of posts for specific interest groups, ensuring content reaches the right audience & Figure \ref{fig:prototypes}(c) \\

\textbf{Intentional Signaling} & Clarifies \textbf{\textit{Ambiguous Reactions}} [Section \ref{section:4-2-4}] by encouraging more meaningful, context-specific reactions beyond simple `Likes' & Figure \ref{fig:prototypes}(f) \\

\textbf{Low-Stakes Disclosure} & Alleviates \textbf{\textit{Presentation Expectations}} [Section \ref{section:4-3-1}] by offering casual, subtle, or ephemeral sharing options that lower the pressure to curate perfect content & Figure \ref{fig:prototypes}(b) \\

\textbf{Contextual Clarity} & Reduces the \textbf{\textit{Social Risk of Misrepresentation}} [Section \ref{section:4-3-2}] common in CMC by allowing users to add contextual information, reducing likelihood of misunderstandings and misjudgments & Figure \ref{fig:prototypes}(e) \\

\textbf{Self-Contained Disclosure} & Prevents \textbf{\textit{Nonconsensual Exposure}} [Section \ref{section:4-3-3}] by ensuring that every user has a dedicated space from the start, enabling them to share without worrying about burdening others by ``clogging'' their feeds & Figure \ref{fig:prototypes}(d) \\

\textbf{Trust-Centered Norms} & Tackles the \textbf{\textit{Cycle of Distrust}} [Section \ref{section:4-3-4}] and cultivates positive community norms by setting clear behavioral guidelines, enabling users with aligned intentions to join and mutually enforce these norms & Figure \ref{fig:prototypes}(h) \\

\bottomrule
\end{tabular}
\end{table*}

\twocolumn

\onecolumn
\section{Appendix: Participant Demographic Information}
\label{ref:individual-demographics}
\begin{longtable}{l p{1cm} p{0.5cm} p{0.8cm} p{2.5cm} p{7.3cm} r r r}
\caption{Survey Responses on Social Media Sharing Concerns} \\
\label{tab:demographics-individual} \\
\toprule
 \textbf{PID} & \textbf{Gender} & \textbf{Age} & \textbf{Race} & \textbf{Social Media Use} & \textbf{Concerns When Sharing on Social Media} & \textbf{**} & \textbf{***} & \textbf{****} \\
\midrule
\endfirsthead

\multicolumn{9}{c}{{\textit{Continued from previous page}}} \\
\toprule
 \textbf{PID} & \textbf{Gender} & \textbf{Age} & \textbf{Race} & \textbf{Social Media Use} & \textbf{Concerns When Sharing on Social Media} & \textbf{**} & \textbf{***} & \textbf{****} \\
\midrule
\endhead

\bottomrule
\multicolumn{9}{r}{{\textit{Continued on next page}}} \\
\endfoot

\bottomrule
\endlastfoot

 P01 & Girl & 16 & White & Instagram, Snapchat, BeReal, Twitter, TikTok & I'm worried I will be judged or attacked for my opinion no matter how passive or considerate I am. & 39 & 27 & 20 \\
 P02 & Boy & 18 & White & Instagram, Snapchat, BeReal, Twitter, TikTok & I'm concerned about my digital footprint being available for all & 47 & 35 & 16 \\
 P03 & Girl & 17 & Asian & Instagram, Snapchat, BeReal, Twitter, TikTok & I am concerned that people may find me bland or they don't care about what I am interested in. I also sometimes worry that my posts won't get as many likes as previous posts. & 42 & 27 & 17 \\
 P04 & Boy & 16 & White & Instagram,\newline{}Twitter & Sharing to much personal information. & 41 & 24 & 13 \\
 P05 & Girl & 16 & Asian & Instagram, Snapchat, BeReal, TikTok & One on my biggest concerns is safety leaks. Even with apps  like Snapchat it is so easy for your posts to get in the wrong hands. As a young teen I also often feel worried about creepy people online. Lastly, it is hard when there is people I may not want to allow to follow me, but I feel pressured to do so in order to seem cooler and gain more followers. & 41 & 32 & 17 \\
 P06 & Girl & 18 & Asian & Instagram, Snapchat, BeReal, Twitter & People that I know in real life will read it and judge me/treat me differently for what I share, or look down on me for different reasons & 35 & 25 & 23 \\
 P07 & Boy & 14 & White & Instagram, Snapchat, BeReal, Twitter & My biggest concerns or reservation when I share about myself on social media is how people will think of myself and how I look. I have started to make my accounts private so only people I approve can see my posts.  & 45 & 29 & 12 \\
 P08 & Girl & 14 & White & Instagram, Snapchat, TikTok & That what I share will be used against me or will be seen by the wrong crowd of people and will be used in malicious ways. & 21 & 23 & 23 \\
 P09 & Girl & 13 & Black & Instagram, Snapchat, BeReal, Twitter, TikTok & People judging me and saying stuff about me & 45 & 38 & 17 \\
 P10 & Girl & 17 & Black & Instagram, Snapchat, BeReal, Twitter, TikTok & I worry about my privacy and how people might react to the information I share. & 41 & 30 & 16 \\
 P11 & NB* & 17 & White & Instagram, Snapchat, BeReal, Twitter, TikTok, Tumblr, Discord, Reddit & That I’m over sharing or that the people I’m sharing with don’t care. And the need to somehow maximize my life experiences to seem as interesting as possible. & 47 & 34 & 16 \\
 P12 & Boy & 15 & White & Instagram, Snapchat, BeReal, Twitter, TikTok & Worried about future image and people I don’t want accessing my posts seeing them & 43 & 28 & 23 \\
 P13 & Girl & 15 & White & Instagram, Snapchat, BeReal, TikTok & Creepy people that are stalking me & 46 & 33 & 22 \\
 P14 & Girl & 13 & White, Asian & Instagram, Snapchat, BeReal, Twitter & My main concern for sharing about myself on social media is privacy. Especially in the case of people with malicious motivations gaining access to my accounts, I sometimes feel nervous about posting, even on private accounts. & 45 & 34 & 15 \\
 P15 & NB* & 17 & White & Instagram, Snapchat, BeReal, Twitter, TikTok & my biggest concern is that people who bullied me in high school see my posts and circulate them and make fun of me. my account has to be public for professional purposes, but i still get scared & 50 & 40 & 18 \\
 P16 & Girl & 15 & Asian & Instagram, BeReal, Reddit, Discord, YouTube & I am kind of anxious that the post won't be "right" so to speak. When I comment, it's normally about something I am passionate about or something that is personal, so I like to keep anonymity (I am still always very respectful regardless.) Many people use editing, and it feels difficult to compete with that and their perfect captions. & 40 & 28 & 20 \\
 P17 & Girl & 16 & White & Instagram,\newline{}BeReal & looking bad, looking too fake, worrying about who is seeing it and what they’re thinking about it, digital footprint, posting too much, etc & 47 & 38 & 15 \\
 P18 & Girl & 18 & Asian & Instagram,\newline{}BeReal & Privacy is my biggest concern. I think it’s mainly the worry about what’s being viewed by who, and the fear that what is shared will stay forever/can be shared to others without my knowledge. I understand that there’s always going to be a lack of privacy when posting in public spaces, but I’d prefer to know if anyone had screenshotted/shared what I have posted. & 42 & 35 & 20 \\
 P19 & Boy & 18 & Asian & Instagram, Twitter, TikTok & Among my biggest concerns with sharing things on social media is the possibility that others may not perceive me the way I want or believe them to. When I post anything, I make sure that nothing about it is embarrassing and that it gives the people who view it the impression that I am mature and self-aware.\newline{}\newline{}Above all, social media has become an important factor in how people present themselves. One of the first things that people explore about someone they meet is their social media profiles; these have become almost essential, for most, as a presentation of their identity. & 43 & 32 & 20 \\
\end{longtable}
\vspace{-5mm}
\noindent{\small\textit{{*}: Non-binary or third gender\\{**}: Peer Trust Score (e.g., ``My friends listen to what I have to say'')\\{***}: Peer Communication Score (e.g., ``When we discuss things, my friends care about my point of view'')\\{****}: Peer Alienation Score (e.g., ``I feel angry with my friends'')}}






\end{document}